\documentclass{article}

\title{Renormalization of lattice-regularized quantum gravity models \\I. General considerations}

\author{Joshua H. Cooperman \\ \emph{Institute for Mathematics, Astrophysics, and Particle Physics}\\ \emph{Radboud Universiteit Nijmegen, Heyendaalseweg 135, 6526 AJ Nijmegen, Nederland}}

\usepackage{float}
\usepackage{multicol}
\usepackage{subfigure}
\usepackage{tabularx}
\usepackage{booktabs}
\usepackage{graphicx}
\usepackage{amsmath}
\usepackage{amsfonts}
\usepackage{latexsym}
\usepackage{mathrsfs}
\usepackage{geometry}
\usepackage{chngcntr}
\counterwithin{figure}{section}
\counterwithin{table}{section}
\numberwithin{equation}{section}

\begin{document}

\maketitle

\begin{abstract}
Lattice regularization is a standard technique for the nonperturbative definition of a quantum theory of fields. Several approaches to the construction of a quantum theory of gravity % as a quantum theory of fields 
adopt this technique either explicitly or implicitly. % and several approaches to the construction of a quantum theory of gravity not as a quantum theory of fields employ what essentially amount to a lattice regularization though it does not have that interpretation. Although one introduces the lattice regularization to define the theory, one is in fact interested in its hypothetical continuum limit. 
%Integral complementing lattice regularization is the process of renormalization. 
A crucial complement to lattice regularization is the process of renormalization through which a continuous description of the quantum theory arises. %by which a continuum description is derived from the lattice-regularized quantum theory 
%one attempts to remove this regularization while physical quantities remain finite. This process requires renormalization. 
I provide a comprehensive conceptual discussion of the renormalization of lattice-regularized quantum gravity models. %, emphasizing concepts over technicalities. 
I begin with a presentation of the renormalization group from the Wilsonian perspective. I then consider the application of the renormalization group in four contexts: quantum field theory on a continuous nondynamical spacetime, quantum field theory on a lattice-regularized nondynamical spacetime, quantum field theory of continuous dynamical spacetime, and quantum field theory of lattice-regularized dynamical spacetime. %Building one upon the next, 
The first three contexts serve to identify successively the particular %successive context to identify the succession of 
issues %and their subtleties 
that arise in the fourth context. These issues originate in the inescability of establishing all scales solely on the basis of the dynamics. While most of this discussion rehearses established knowledge, the attention that I pay to %my emphasis on 
these issues, especially the previously underappreciated role of standard units of measure, is largely novel. 
%but I bring it together in one place with an emphasis on a particular set of issues. My discussion of the role of standard units of measurement is novel. 
I conclude by briefly reviewing past studies of renormalization of lattice-regularized quantum gravity models. In the second paper of this two-part series, I illustrate the ideas presented here by proposing a renormalization group scheme for causal dynamical triangulations. 

%The aim of the causal dynamical triangulations approach is to construct a quantum theory of gravity as the continuum limit of a lattice-regularized model of dynamical geometry. A renormalization group scheme---in concert with finite size scaling analysis---is essential to this aim.  Formulating and implementing such a scheme in this context raises novel and notable conceptual and technical problems. %, however, is notably nontrivial in comparison to its implementation within a lattice-regularized model defined on a nondynamical geometry. 
%I first explore and then resolve % the conceptual and technical problems facing a renormalization group scheme for dynamical geometry, and I advance  
%these problems. As an application of my solution, I propose a definite renormalization group scheme for causal dynamical triangulations. 
\end{abstract}

\section{Introduction}\label{introduction}

The renormalization group forms the foundation of the modern understanding of quantum theories of fields. If the quantum theory of gravity %---or just a quantum theory of gravity---
proves to be yet another quantum theory of fields, %proves to be yet another triumph of the field theory paradigm, 
then the renormalization group should also underpin %an understanding of 
this much sought-after theory.%the quantum theory of gravity.
\footnote{The renormalization group might underpin the quantum theory of gravity even if it is not a quantum theory of fields.} Several contemporary approaches attempt to construct candidate quantum theories of gravity as quantum theories of fields. For instance, within the asymptotic safety program one employs functional renormalization group techniques in an attempt to base a quantum theory of gravity on an ultraviolet non-Gaussian fixed point \cite{MN&MR}, and within the causal dynamical triangulations program one employs lattice regularization techniques in an attempt to define a quantum theory of gravity in the continuum limit \cite{JA&AG&JJ&RL3}. Of course, whether or not the quantum theory of gravity is a quantum theory of fields remains far from determined. 

To inform this determination from a theoretical perspective, one should understand these approaches within the context of the renormalization group. Notoriously, nonperturbative approaches to the construction of quantum theories of gravity, like the asymptotic safety and causal dynamical triangulations programs, %particularly those based on a lattice regularization, 
present notable challenges to the achievement of such an understanding. %to the formulation and implementation of a renormalization group scheme. 
These challenges all stem from one indubitable fact: all length scales are dynamically determined in a theory of gravity. The renormalization group is fundamentally concerned with the relation between theories describing the same physics yet effective on different length scales. Thus, even to begin to understand these approaches within the context of the renormalization group, one requires a handle on the quantum dynamical emergence of length scales. This is no small task. 

The asymptotic safety program nevertheless puts the renormalization group front and center: its practitioners seek the desired fixed point by studying % study the renormalization group flows solving the Wetterich equation in search of the desired fixed point. 
the renormalization group flows that solve the Wetterich equation for gravity \cite{MR,CW}. The formulation of this exact renormalization group equation relies on the adoption of a particular hypothesis regarding the establishment of length scales \cite{MR}. 
%its practitioners have necessarily attempted to understand the quantum theory of gravity so defined within the context of the renormalization group. Its practitioners adopt a particular \emph{Ansatz}  
I comment only briefly on this hypothesis in the following, postponing a critique to future work. %Although inescapably 
On the contrary, the causal dynamical triangulations program delays an attempt at application of the renormalization group: its practitioners study the phenomenology presented by their approach and only subsequently attempt a renormalization group analysis. %is also inescapably a renormalization group based approach, the application of the renormalization group in this context is much less immediate. 
Before focusing on the causal dynamical triangulations program in the second paper of this two-part series \cite{JHC}, to which I hereafter refer as paper II, I wish to establish clearly the meaning and formulate systematically the process of a renormalization group analysis of lattice-regularized quantum gravity models. This is the primary purpose of this first paper. %Whether or not the techniques of renormalization even apply to a quantum theory of gravity remains to be determined, but for the moment I will assume that the prospective quantum theory of gravity is essentially a quantum field theory to which one can apply the techniques of renormalization. As I will explain below, 
I begin in subsection \ref{conceptualRG} by painting a conceptual picture of the renormalization group from the Wilsonian perspective. I then illustrate this picture in subsection \ref{concreteRG} by considering four cases: quantum field theory on a continuous nondynamical spacetime, quantum field theory on a lattice-regularized nondynamical spacetime, quantum field theory of continuous dynamical spacetime, and quantum field theory of lattice-regularized dynamical spacetime. I intend the first three cases to illuminate successively the elements---and their subtleties---required for the formulation and implementation of a renormalization group scheme in the fourth case. %progressively illuminating the subtleties the arise in the case at hand.
I sketch the experimental measurement of renormalization group flows in subsection \ref{experiment}, showing how closely it parallels the procedure established in subsection \ref{concreteRG}. %To demonstrate the plausibility of my account, in subsection \ref{experiment} I compare it to how experimental observations of renormalization group flows are actually performed. 
I review previous renormalization group studies of lattice-regularized quantum gravity models in section \ref{review} before concluding in section \ref{conclusion}.
%so the informed reader may wish to skip to section \ref{theory}. %but I wish to establish clearly all of the necessary ingredients for the definition of the renormalization group.

Throughout this paper I am principally concerned with establishing a conceptual understanding of the renormalization group in the context of lattice-regularized quantum gravity models. Accordingly, I eschew technical details for the sake of conceptual clarity, and I keep inessential details at bay by maintaining a high degree of generality. In paper II I am principally concerned with devising a renormalization group scheme for causal dynamical triangulations, which also serves the purpose of providing a concrete example illustrating in complete detail all of the ideas presented in this paper. 
%my secondary purpose, beyond the primary purpose of contributing to an understanding of renormalization for causal dynamical triangulations, is to provide a detailed and concrete example illustrating the ideas discussed in this paper. % reserve technical details for paper II in which I illustrate the concepts elaborated here with a definite example. 
Much of my discussion, particularly that of subsection \ref{conceptualRG} and the first two subsubsections of subsection \ref{concreteRG}, largely follows standard presentations, for instance, those of \cite{JC,NG,MEP&DVS,HJR}. The novelty of my discussion stems from its emphasis on the assumptions that one makes in standard presentations which are no longer applicable in the context of lattice-regularized quantum gravity models. As I asserted above, the absence of a means to define length scales, including a standard unit of length, independently of the dynamics invalidates these assumptions.
%most of these assumptions are invalidated by the lack of a means to define length scales independently of the quantum dynamics, both a standard unit of length, which I emphasize especially, and the establishment of dynamical scales in units of a standard. %My emphasis on the subtleties arising in the present context is to some extent novel. I also stress the role of a standard unit of measure. %Much of my discussion appears in many disparate sources. I attempt to provide a compact synthesis, filling in the gaps that arise in the present context. Much of what I present appears elsewhere in the literature, but I do not know of a single comprehensive presentation for the context of quantum gravity models.
%I aim to bring together the insights of this literature with my own. %This is certainly the case for the two subsubsections of subsection \ref{concreteRG}. Most of what I present in the last two subsubsections is also reasonably well known, though I do not know of a presentation that puts together all of the pieces. Even the studies of renormalization that I review in section \ref{review} do not contain complete discussions. 
Although the literature contains several discussions of renormalization of lattice-regularized quantum gravity models---most notably \cite{JA&AG&JJ&AK&RL,PB&ZB&AK&BP,JH,MN} in my opinion---I do not know of a single presentation that comprehensively discusses all of the attendant issues. By bringing together the insights of this literature with my own, I aim to fill this gap. While I focus on lattice-regularized quantum gravity models, from which one ultimately wishes to extract a continuous description, I hope that my discussion proves relevant for other types of discrete quantum gravity models. Since a description of gravity in terms of continuous quantities works exceptionally well on sufficiently large scales, any discrete quantum gravity model must yield such a description, and application of the renormalization group might greatly assist in its derivation. %on the appropriate interval of scales. 

%\subsection{Preview of paper II}

\section{A renormalization group primer}\label{defRGflows}

\subsection{Conceptualization of the renormalization group}\label{conceptualRG}

\subsubsection{Organizing theory space}
%\subsubsection{Functioning}

Consider the space $\mathfrak{T}$ of all theories $\mathscr{T}$ of a particular set of dynamical fields $\Phi$. Suppose that each such theory $\mathscr{T}$ is specified by a Lagrangian
\begin{equation}\label{lagrangian}
\mathscr{L}_{\mathscr{T}}[\Phi]=\sum_{j}c_{j}f_{j}[\Phi],
\end{equation}
a linear combination of scalar functionals $f_{j}[\Phi]$ of the fields $\Phi$. Each scalar functional $f_{j}[\Phi]$ describes a particular interaction of the fields $\Phi$, and each coefficient $c_{j}$ gives the coupling for that interaction. One may thus parametrize the space $\mathfrak{T}$ by the couplings $c_{j}$, a set of values for the couplings $c_{j}$ specifying a theory $\mathscr{T}$.\footnote{I ignore the possibility of field redefinitions rendering the set of all scalar functionals $f_{j}[\Phi]$ overcomplete. Field redefinitions lead of course to coupling redefinitions, which just consist of reparametrizations of the space $\mathfrak{T}$. Ultimately, one defines the physical meaning of a coupling $c_{j}$ through a procedure for measuring its value.} One often chooses to impose certain physically motivated restrictions on the set of all possible interactions, implemented by allowing only a subset of scalar functionals to contribute to the Lagrangian \eqref{lagrangian}. %From the Lagrangian \eqref{lagrangian} one constructs the quantum effective action as
%\begin{equation}
%S[\Phi]=\int_{\mathcal{M}}\mathrm{d}^{4}x
%\end{equation}
One such subset constitutes a truncation $\mathfrak{t}$ of the space $\mathfrak{T}$. Typically, one obtains a truncation $\mathfrak{t}$ (in part) by restricting to scalar functionals $f_{j}[\Phi]$ that are invariant under the action of certain symmetry transformations. Even with reasonable restrictions on the set of all possible scalar functionals, a truncation $\mathfrak{t}$ of the space $\mathfrak{T}$ is typically prodigious. How does one then make sense of the space $\mathfrak{T}$?

The renormalization group serves to organize the space $\mathfrak{T}$. The key to this organization is the following realization: each theory $\mathscr{T}$ provides a description of the physics of the fields $\Phi$ effective only within an interval of length scales $(\ell_{\mathrm{UV}},\ell_{\mathrm{IR}})$. %\footnote{Whenever I refer to a scale, I refer to a length scale.} 
The ultraviolet scale $\ell_{\mathrm{UV}}$ is of course smaller than the infrared scale $\ell_{\mathrm{IR}}$; these scales might coincide in which case the theory $\mathscr{T}$ is effective at this common scale. One should not conceive of the scale $\ell_{\mathrm{UV}}$ as an ultraviolet regulator and the scale $\ell_{\mathrm{IR}}$ as an infrared regulator for some regularization of, for instance, a path integral for the fields $\Phi$. The scales $\ell_{\mathrm{UV}}$ and $\ell_{\mathrm{IR}}$ are rather the physical scales delimiting the regime of validity of the theory $\mathscr{T}$. I introduce regularizing scales explicitly in subsection \ref{concreteRG}, discussing there the interplay of regularization and renormalization. The interval $(\ell_{\mathrm{UV}},\ell_{\mathrm{IR}})$ is a range of scales that one can probe experimentally given the conditions of one's experiment, and the theory $\mathscr{T}$ is that which provides a description of one's experimental results on this range of scales. 

Not every theory $\mathscr{T}$, however, necessarily provides a description of the same physics of the fields $\Phi$. %: some subspace of theories $\mathscr{T}$  effective within a different interval of length scales $(\ell_{\mathrm{UV}},\ell_{\mathrm{IR}})$. 
The renormalization group sorts the space $\mathfrak{T}$ into subspaces $\mathfrak{T}_{l}$, the theories $\mathscr{T}^{(l)}$ in each of which describe the same physics within different (but not necessarily mutually exclusive) intervals of scales. The renormalization group moreover relates the theories $\mathscr{T}^{(l)}$ within the subspace $\mathfrak{T}_{l}$: a renormalization group transformation connects a theory $\mathscr{T}^{(l)}$ valid within the interval $(\ell_{\mathrm{UV}},\ell_{\mathrm{IR}})$ to a theory $\mathscr{T'}^{(l)}$ valid within the interval $(\ell_{\mathrm{UV}}',\ell_{\mathrm{IR}}')$. In particular, a renormalization group transformation maps the couplings $c_{j}$ specifying the theory $\mathscr{T}^{(l)}$ into the couplings $c_{j}'$ specifying the theory $\mathscr{T'}^{(l)}$. The couplings $c_{j}$ transform into the couplings $c_{j}'$ in precisely such a manner that the theories $\mathscr{T}^{(l)}$ and $\mathscr{T'}^{(l)}$ describe the same physics  within their respective intervals of scales. The physical necessity of this transformation of the couplings is evident: %must change under the action of a renormalization group transformation because 
for consistency with the theory $\mathscr{T}^{(l)}$, the theory $\mathscr{T'}^{(l)}$ must encode the effects of the physics occurring within the interval $(\ell_{\mathrm{UV}},\ell_{\mathrm{IR}})\setminus (\ell_{\mathrm{UV}}',\ell_{\mathrm{IR}}')$. Indeed, if one uses the theory $\mathscr{T}^{(l)}$ to compute the physical observable $\mathscr{O}(\ell)$, and one uses the theory $\mathscr{T'}^{(l)}$ to compute the physical observable $\mathscr{O}(\ell)$, then one obtains the same predictions for $\mathscr{O}(\ell)$ within the interval $(\ell_{\mathrm{UV}},\ell_{\mathrm{IR}})\cap(\ell_{\mathrm{UV}}',\ell_{\mathrm{IR}}')$, those length scales $\ell$ on which both theories are valid. Since the values of a complete set of physical observables $\mathscr{O}(\ell)$ capture the physics of a theory $\mathscr{T}$, the previous sentence conveys the precise sense in which two theories $\mathscr{T}^{(l)}$ and $\mathscr{T}'^{(l)}$ describe the same physics. By incrementing a renormalization group transformation, one generates the corresponding renormalization group trajectory, a sequential ordering of the theories $\mathscr{T}^{(l)}$. Along a renormalization group trajectory, the couplings change incrementally, each tracing out a renormalization group flow. 

%The reason for this consistency is simple: renormalization group transformation do not change the dynamics of the fields $\Phi$; rather, renormalization group transformation encode the dynamics within the interval $(\ell_{\mathrm{UV}},\ell_{\mathrm{IR}})\setminus(\ell_{\mathrm{UV}}',\ell_{\mathrm{IR}}')$ in the couplings $c'_{j}$. 

Following standard practice, I employ the term `theory' %has (at least) two related but distinct connotations. As I have defined a theory $\mathscr{T}$ in subsection \ref{conceptualRG}, it is specified by a Lagrangian $\mathscr{L}_{\mathscr{T}}[\Phi]$ for fixed values of the couplings $c_{j}$. One often defines a theory as consisting of an entire renormalization group trajectory, encompassing all of the theories $\mathscr{T}$ along the trajectory. When I refer to a theory with a label $\mathscr{T}$, I intend the first connotation. When I refer to a theory without a label, I intend the second connotation.} 
also to designate an entire renormalization group trajectory. To distinguish the two connotations of the term `theory'---the former signifying that which is specified by a Lagrangian $\mathscr{L}_{\mathscr{T}}[\Phi]$ for fixed values of the couplings $c_{j}$, the latter signifying that which is specified by all of the Lagrangians $\mathscr{L}_{\mathscr{T}}[\Phi]$ along a renormalization group trajectory---I adopt the following convention. When I refer to a theory with a label $\mathscr{T}$, I intend the first connotation, and when I refer to a theory without a label, I intend the second connotation. One often conceives of the couplings $c_{j}$ as functions of the interval of scales $(\ell_{\mathrm{UV}},\ell_{\mathrm{IR}})$. These functions---the couplings' renormalization group flows---then specify a theory (in the second connotation). 

A renormalization group transformation generically changes both the ultraviolet scale $\ell_{\mathrm{UV}}$ and the infrared scale $\ell_{\mathrm{IR}}$. %, and the intervals $(\ell_{\mathrm{UV}},\ell_{\mathrm{IR}})$ and $(\ell_{\mathrm{UV}}',\ell_{\mathrm{IR}}')$ certainly need not be disjoint. 
In the most widely considered case, the renormalization group transformation incrementally increases the ultraviolet scale $\ell_{\mathrm{UV}}$ to $\ell_{\mathrm{UV}}'=\ell_{\mathrm{UV}}+\delta\ell$ and leaves fixed the infrared scale $\ell_{\mathrm{IR}}$ (often assumed to be infinite). Such a transformation constitutes a coarse graining of the degrees of freedom of the fields $\Phi$. In this case the renormalization group tracks how an initial theory $\mathscr{T}^{(l)}$ changes as the dynamics of the fields $\Phi$ on small scales are successively integrated out. While many discussions of renormalization are couched within the context of a coarse graining operation, one can formulate renormalization group equations, such as the Polchinski and Wetterich equations, that describe continuous changes in both of the scales $\ell_{\mathrm{UV}}$ and $\ell_{\mathrm{IR}}$ \cite{JP,CW}. In general, one tailors a renormalization group transformation to change the ultraviolet scale $\ell_{\mathrm{UV}}$ and the infrared scale $\ell_{\mathrm{IR}}$ in accordance with the scales that one wishes to probe. %given some initial condition, a particular theory $\mathscr{T}$.}

\subsubsection{Classifying fixed points}

In addition to organizing the space $\mathfrak{T}$ into the subspaces $\mathfrak{T}_{l}$ and ordering the theories $\mathscr{T}^{(l)}$ within each subspace $\mathfrak{T}_{l}$, the renormalization group induces further physically significant structure. Certain theories $\mathscr{T}_{*}$ are identified as fixed points of the renormalization group transformation: their couplings $c_{j}^{*}$ do not change under its action. Relative to a fixed point theory $\mathscr{T}_{*}$, each scalar functional $f_{j}[\Phi]$ is classified as relevant, marginal, or irrelevant depending on how its associated coupling $c_{j}$ changes in the vicinity of that fixed point theory $\mathscr{T}_{*}$.\footnote{Technically, one linearizes the renormalization group transformation in a sufficiently small neighborhood of the fixed point theory $\mathscr{T}_{*}$. Treating this transformation as a linear map acting on the (differences of the) couplings $c_{j}$ (from their fixed point values $c_{j}^{*}$), one determines the linear combinations of couplings that form eigenvectors of this map. The associated linear combinations of the scalar functionals $f_{j}[\Phi]$ are then classified as relevant, marginal, or irrelevant according to whether their associated eigenvalues increase, remain constant, or decrease under a linearized renormalization group transformation.} Specifically, near a fixed point theory $\mathscr{T}_{*}$, under a renormalization group transformation, a coupling driven away from its value at the fixed point characterizes a relevant scalar functional, a coupling neither driven away from nor towards its value at the fixed point characterizes a marginal scalar functional, and a coupling driven towards its value (of zero) at the fixed point characterizes an irrelevant scalar functional. Relevant scalar functionals are relevant in the sense that the interactions which they describe yield the dominant contributions to the dynamics of the fields $\Phi$ on the interval of scales controlled by the fixed point. Marginal scalar functionals are usually either marginally relevant or marginally irrelevant, a distinction only revealed by studying an extended vicinity of the fixed point. A scalar functional may be exactly marginal in which case its associated coupling does not change along renormalization group flows. Irrelevant scalar functionals are irrelevant in the sense that the interactions which they describe yield subdominant contributions to the dynamics of the fields $\Phi$ on the interval of scales controlled by the fixed point. %These subdominant contributions vanish at the fixed point.

Each relevant scalar functional defines an unstable or repulsive direction of the fixed point under renormalization group transformations, and each irrelevant scalar functional defines a stable or attractive direction of the fixed point under renormalization group transformations. If a fixed point has only stable or only unstable directions, then the fixed point itself is deemed stable or unstable. %(in the sense just explained that a small perturbation is driven back to or away from the fixed point)
The subspace $\mathfrak{T}_{c}^{(*)}$ of theories $\mathscr{T}_{c}^{(*)}$ spanned by a fixed point's stable directions---or, equivalently, the subspace of theories on renormalization group trajectories that flow into the fixed point under successive renormalization group transformations---is called the critical manifold.
%The subspace of %the space $\mathfrak{T}$ 
%of theories on renormalization group trajectories that flow into the fixed point under successive renormalization group transformations is called the critical manifold. 
From the definitions of relevance and irrelevance, one readily concludes that the critical manifold is parametrized by the couplings of irrelevant scalar functionals for values of the couplings of relevant scalar functionals tuned to those at the fixed point. %is the critical subspace of the fixed point theory $\mathscr{T}_{*}$. 
%From any theory $\mathscr{T}$ within the critical subspace, renormalization group trajectories flow into the fixed point. 
%All of the irrelevant scalar functionals are thus irrelevant to the physics described by the fixed point theory. Each relevant scalar functional defines a direction in the space $\mathfrak{T}$ along which renormalization group trajectories move away from the fixed point. 
All of the theories $\mathscr{T}_{c}^{(*)}$ within the critical manifold belong to the same universality class: aside from subdominant physical effects, which vanish at the fixed point, the physics of all of these theories is the same or universal. 

Fixed points themselves are classified in various terms. %The values of the couplings $c_{j}^{*}$ of a fixed point theory $\mathscr{T}_{*}$ determines its nature. 
If all of the couplings characterizing interactions (as opposed to free propagation) of the fields $\Phi$ vanish at a fixed point, then that fixed point is Gaussian. A theory defined by a renormalization group trajectory including a Gaussian fixed point exhibits asymptotic freedom in its vicinity. At a Gaussian fixed point the only relevant scalar functionals are thus those describing free propagation of the fields $\Phi$, possibly including a scalar functional endowing the fields $\Phi$ with mass. If the couplings characterizing a finite number of interactions of the fields $\Phi$ approach finite values at a fixed point, and if all of the remaining couplings characterizing interactions of the fields $\Phi$ vanish at that fixed point, then that fixed point is non-Gaussian. A theory defined by a renormalization group trajectory including a non-Gaussian fixed point exhibits asymptotic safety in its vicinity. At a non-Gaussian fixed point the scalar functionals corresponding to these finite couplings are all relevant. These are the only known types of fixed points of the renormalization group flow. None of the couplings can diverge at a fixed point since this entails physically divergent quantities.  %the physics on ever smaller scales is incorporated,  One may inquire into the existence of an ultraviolet fixed point. 

One may further classify a fixed point as ultraviolet or infrared. Consider a renormalization group transformation that incrementally decreases the ultraviolet scale $\ell_{\mathrm{UV}}$ and leaves fixed the infrared scale $\ell_{\mathrm{IR}}$ so that one probes ever smaller length scales. A fixed point reached as this renormalization group transformation iterates the ultraviolet scale $\ell_{\mathrm{UV}}$ to zero is an ultraviolet fixed point. A theory defined by a renormalization group trajectory including an ultraviolet fixed point is ultraviolet-complete: the theory $\mathscr{T}_{*}^{(\mathrm{UV})}$ obtained in the limit of vanishing $\ell_{\mathrm{UV}}$---the continuum limit---provides a description in terms of continuous quantities valid at arbitrarily small length scales. Such a theory is also deemed ultraviolet-renormalizable for reasons that I make clear in subsubsection \ref{QFTCNDST}. I have of course assumed that the couplings characterizing relevant scalar functionals are tuned to criticality so that this renormalization group trajectory reaches the fixed point. In the absence of an ultraviolet fixed point, the quantum theory is ultraviolet-incomplete or ultraviolet-nonrenormalizable since it does not apply on arbitrarily small scales. 

Consider now a renormalization group transformation that leaves fixed the ultraviolet scale $\ell_{\mathrm{UV}}$ and incrementally increases the infrared scale $\ell_{\mathrm{IR}}$ so that one probes ever larger length scales. A fixed point reached as this renormalization group transformation iterates the infrared scale $\ell_{\mathrm{IR}}$ to infinity is an infrared fixed point. A theory defined by a renormalization group trajectory including an infrared fixed point is infrared-complete: the theory $\mathscr{T}_{*}^{(\mathrm{IR})}$ obtained in the limit of infinite $\ell_{\mathrm{IR}}$ provides a description in terms of continuous quantities valid at arbitrarily large length scales. Such a theory is also deemed infrared-renormalizable for reasons that I make clear in subsubsection \ref{QFTCNDST}. I have of course assumed that the couplings characterizing relevant scalar functionals are tuned to criticality so that this renormalization group trajectory reaches the fixed point. In the absence of an infrared fixed point, the quantum theory is infrared-incomplete or infrared-nonrenormalizable since it does not apply on arbitrarily large scales. 
%If the associated renormalization group flow approaches a fixed point in the ultraviolet, then one can take the limit of vanishing lattice spacing. If the associated renormalization group flow approaches a fixed point in the infrared, then one can take the limit of infinite lattice extent. If the associated renormalization group flow does not contain one or both of these fixed points, (couplings blow up) then one can likely still obtain a continuum description effective on a nontrivial interval of scales. 
%The couplings of the scalar functionals relevant at this fixed point must be tuned so that the trajectory reaches the fixed point. 
%The presence of a continuum limit depends on the presence of fixed points of the renormalization group flow. 
%This discussion now clarifies the role of the regularization.
%If an ultraviolet fixed point exists, then one can take the limit of vanishing $\mathfrak{l}_{\mathrm{UV}}$, and, if an infrared fixed point exists, then one can take the limit of infinite $\mathfrak{l}_{\mathrm{IR}}$. 

%In these cases the quantum theory is renormalizable. If either or both of these 
In the absence of fixed points, one obtains a nonrenormalizable theory defined by a renormalization group trajectory that terminates neither in the ultraviolet nor in the infrared. This theory may nevertheless prove effective on a nontrivial interval of scales. In fact, there is presumably little empirical difference between renormalizable and nonrenormalizable theories since the latter can mimic the former over a considerable interval of scales.
%a nonrenormalizable quantum theory can mimic a renormalizable quantum theory over a considerable interval of scales. 
As I explained just above, a renormalizable theory is defined by a renormalization group trajectory connecting an ultraviolet to an infrared fixed point. %There is only one renormalization group trajectory connecting these two fixed points. Thus, only the single quantum theory defined by this one trajectory is strictly renormalizable. 
There are infinitely many renormalization group trajectories within the space $\mathfrak{T}$ that are close (in an appropriate measure) to this distinguished renormalization group trajectory. Given some level of empirical ability, specified by the interval of scales $(\ell_{\mathrm{UV}},\ell_{\mathrm{IR}})$ that one can probe, one can always find a renormalization group trajectory defining a nonrenormalizable theory that exhibits physics sufficiently close to that of the renormalizable theory such that the two theories are empirically indistinguishable. %Accordingly, to determine empirically that a renormalizable quantum theory describes some system is presumably not only difficult, but also unnecessary. 
Furthermore, although a renormalizable theory applies on all scales in principle, this theory likely does not apply on all scales in practice because novel physics for which one previously did not account may well enter on either very small or very large scales. A renormalizable theory is just as ignorant of this physics as is a nonrenormalizable theory. 

Identifying the fixed point theories $\mathscr{T}_{*}$ and categorizing the scalar functionals $f_{j}[\Phi]$ are primary motivations for a renormalization group analysis. The presence and properties of the fixed point theories $\mathscr{T}_{*}$ control the structure of the space $\mathfrak{T}$ induced by renormalization group transformations. This structure encodes a considerable amount of information concerning the physics of the fields $\Phi$. %Since I only discuss the workings of a renormalization group scheme here, I do not wish to dwell on these concepts, but I return to them briefly in section \ref{conclusion}.
Owing to the structure deriving from a fixed point, renormalization group trajectories that pass sufficiently close to the fixed point but do not reach the fixed point exhibit physics sufficiently close to that dictated by the fixed point to be largely indistinguishable. One's requisite degree of indistinguishability is dictated by one's purposes, be they experimental or theoretical.

%One is particularly interested in the absence or presence of fixed points of the renormalization group flow. 

%first determine a trajectory of constant physics within the bare coupling space of causal dynamical triangulations, and then perform the technique outlined in the previous paragraph for each point along this trajectory. One must thus address the question of what constitutes such a trajectory.

%To specify a renormalization group scheme, I must first establish how the renormalization group trajectories should be defined. Thus, this first step is really a prerequisite for all renormalization group schemes. 

%Now, suppose that a theory $\mathcal{T}_{0}$ is valid over the interval of scales $(\ell_{\mathrm{UV}},\ell_{\mathrm{IR}})$. In each step of the renormalization group scheme, suppose that one ends up with a theory $\mathcal{T}_{n}$ valid over the interval of scales $(\ell_{\mathrm{UV}}+n\delta\ell,\ell_{\mathrm{IR}})$ for some reasonable small increment $\delta\ell$ and $n\in\{1,\ldots,N\}$. 

\subsubsection{Defining length scales}

Implicit in this discussion is the assumption that the length scales $\ell_{\mathrm{UV}}$ and $\ell_{\mathrm{IR}}$---and all of those in between---have a well-defined physical meaning. Integral to giving these length scales a well-defined physical meaning is the establishment of a standard unit of length $\ell_{\mathrm{unit}}$ in terms of which these length scales are measured.\footnote{Indeed, the physical meaning of a dimensionful quantity is that one can only ascertain its value in units of a chosen standard for the relevant dimensions.} %In the case of length scales, one can only distinguish small scales, like $\ell_{\mathrm{UV}}$, from large scales, like $\ell_{\mathrm{IR}}$,  one checks that the ratio $\ell_{\mathrm{UV}}/\ell_{\mathrm{unit}}<\ell_{\mathrm{IR}}/\ell_{\mathrm{unit}}$.} 
As I intend the four cases of subsection \ref{concreteRG} to demonstrate, the definition of the scales $\ell_{\mathrm{UV}}$, $\ell_{\mathrm{IR}}$, and $\ell_{\mathrm{unit}}$---by which they acquire a well-defined physical meaning---is either independent of or dependent on the dynamics of the theory under consideration. In the case of independence from the dynamics, one first defines a standard unit of length $\ell_{\mathrm{unit}}$ and a means for measuring other length scales in terms of $\ell_{\mathrm{unit}}$ within a separate physical model, and one then imports this model as a fixed background structure into the theory. Independence of these scales from the dynamics is well-justified only in certain restricted settings, specifically those in which the dynamics of the theory enter neither into the establishment of a standard unit of length nor into the determination of other length scales (measured in terms of this unit). %Dependence of these scales on the dynamics If this assumption is not justified---as one might argue is strictly always the case---then one must employ an appropriate alternative.
In the case of dependence on the dynamics, one must determine how to define a standard unit of length $\ell_{\mathrm{unit}}$ and a means for measuring other length scales in terms of this unit from the dynamics itself. Such a determination might prove quite nontrivial since one must identify physical observables that define the relevant length scales. %, and one must have sufficient physical insight to grasp that these physical observable do indeed define length scales. 
In actuality, as I intend the discussion of subsection \ref{experiment} to emphasize, one does always define physically meaningful length scales on the basis of dynamical phenomena. The distinction between independence and dependence of scales on the dynamics is then recast as the degree to which the dynamics defining a standard unit of length $\ell_{\mathrm{unit}}$ and the dynamics correlating a standard unit of length $\ell_{\mathrm{unit}}$ with other length scales %and correlating the standard unit of length with other length scales
are unaffected by the remaining dynamics. %This is also the theoretical setting that has been most applied to actual measurements. Let us briefly consider how these measurements proceed. In particular, how do measurements yield the renormalization group flow of a coupling? 
%With this conceptual picture of the renormalization group in place, I now consider four successive examples that illustrate this picture, its particulars and its nuances. 

\subsection{Concretization of the renormalization group}\label{concreteRG}

\subsubsection{Quantum field theory on a continuous nondynamical spacetime}\label{QFTCNDST}

\paragraph{Definition}

Consider first the quantum theory of the field $\Phi$ propagating on a fixed spacetime manifold $\mathscr{M}$ with metric tensor $\mathbf{g}$.
Assume that the Lagrangian $\mathscr{L}_{\mathscr{T}}[\Phi]$ provides an effective description of the dynamics within the interval of scales $(\ell_{\mathrm{UV}},\ell_{\mathrm{IR}})$. How might one arrive at the Lagrangian $\mathscr{L}_{\mathscr{T}}[\Phi]$? Since I am concerned in paper II with the quantization of a classical theory by path integral techniques, I suppose that one obtains the Lagrangian $\mathscr{L}_{\mathscr{T}}[\Phi]$ by such a method. The conceptual content of my account should nevertheless apply to other techniques for computing the Lagrangian $\mathscr{L}_{\mathscr{T}}[\Phi]$.

Accordingly, one starts from a classical theory $\mathscr{T}_{\mathrm{cl}}$ of the field $\Phi$ specified by the Lagrangian $\mathscr{L}_{\mathrm{cl}}[\Phi]$. Working in a coordinate representation of the spacetime manifold $\mathscr{M}$, one introduces the classical action 
\begin{equation}
S_{\mathrm{cl}}[\Phi]=\int_{\mathscr{M}}\mathrm{d}^{d+1}x\sqrt{-g}\mathscr{L}_{\mathrm{cl}}[\Phi].
\end{equation}
One then formally computes a transition amplitude $\mathscr{A}[\psi]$ in the quantum theory as the path integral
\begin{equation}\label{QFT1state}
\mathscr{A}[\psi]=\int_{\Phi|_{\partial\mathscr{M}}=\psi}\mathrm{d}\mu(\Phi)\,e^{iS_{\mathrm{cl}}[\Phi]/\hbar}.
\end{equation}
The configuration $\psi$ of the field $\Phi$ on the boundary $\partial\mathscr{M}$ of the spacetime manifold $\mathscr{M}$ specifies the transition amplitude $\mathscr{A}[\psi]$. The expression \eqref{QFT1state} is an instruction to integrate over all configurations of the field $\Phi$ satisfying the boundary condition $\Phi|_{\partial\mathscr{M}}=\psi$, weighting each configuration by the product of the measure $\mathrm{d}\mu(\Phi)$ and the exponential $e^{iS_{\mathrm{cl}}[\Phi]/\hbar}$. There is some freedom in the definition of the measure $\mathrm{d}\mu(\Phi)$. %At the very least one 
One usually designs the measure $\mathrm{d}\mu(\Phi)$ to respect any invariances of the theory $\mathscr{T}_{\mathrm{cl}}$ so that one does not introduce additional degrees of freedom and to eliminate any redundancies of description of the theory $\mathscr{T}_{\mathrm{cl}}$ so that one integrates only over physically distinct configurations of the field $\Phi$. In certain cases one may not be able to achieve one or both of these aims.

The path integral in equation \eqref{QFT1state} is often formally divergent, so one must introduce a regularization to render it well-defined. Typically, one requires an ultraviolet regulator $\mathfrak{l}_{\mathrm{UV}}$ to prevent excitations of the field $\Phi$ on the smallest scales from contributing to the path integration, and, possibly, one requires an infrared regulator $\mathfrak{l}_{\mathrm{IR}}$ to prevent excitations of the field $\Phi$ on the largest scales from contributing to the path integration. %In the absence of these regulators, the path integration in equation \eqref{QFT1state} yields a formally divergent result. 
To enforce these restrictions on the path integration in equation \eqref{QFT1state}, one employs a regularized measure $\mathrm{d}\mu_{(\mathfrak{l}_{\mathrm{UV}},\mathfrak{l}_{\mathrm{IR}})}(\Phi)$ designed to assign zero weight to excitations of the field $\Phi$ on scales $\ell\in(0,\mathfrak{l}_{\mathrm{UV}})\cup(\mathfrak{l}_{\mathrm{IR}},\infty)$. (In the formal expression \eqref{QFT1state}, $\mathfrak{l}_{\mathrm{UV}}=0$ and $\mathfrak{l}_{\mathrm{IR}}=\infty$.) A conceptually straightforward choice for this measure, which does not necessarily respect the invariances of the theory $\mathscr{T}_{\mathrm{cl}}$, is the following:
\begin{equation}\label{QFT1measure1}
\mathrm{d}\mu_{(\mathfrak{l}_{\mathrm{UV}},\mathfrak{l}_{\mathrm{IR}})}(\Phi)=\mathscr{N}\prod_{\ell\in(\mathfrak{l}_{\mathrm{UV}},\mathfrak{l}_{\mathrm{IR}})}\mathrm{d}\phi_{\ell},
\end{equation}
where $\mathscr{N}$ is a normalization factor, and $\phi_{\ell}$ denotes a mode of the field $\Phi$ of length scale $\ell$. %One now performs 
The path integration in equation \eqref{QFT1state} now extends only over those modes $\phi_{\ell}$ for which the scale $\ell\in(\mathfrak{l}_{\mathrm{UV}},\mathfrak{l}_{\mathrm{IR}})$. %Those modes $\phi_{\ell}$ for which the scale $\ell\in(0,\mathfrak{l}_{\mathrm{IR}})\cup(\mathfrak{l}_{\mathrm{IR}},\infty)$ are assigned zero weight. 
Computed in the regularized quantum state 
\begin{equation}\label{QFT1state2}
\bar{\mathscr{A}}[\psi]=\int_{\Phi|_{\partial\mathscr{M}}=\psi}\mathrm{d}\mu_{(\mathfrak{l}_{\mathrm{UV}},\mathfrak{l}_{\mathrm{IR}})}(\Phi)\,e^{iS_{\mathrm{cl}}[\Phi]/\hbar},
\end{equation}
the expectation value $\mathbb{E}_{\bar{\mathscr{A}}[\psi]}[\mathscr{O}]$ of a physical observable $\mathscr{O}$,
\begin{equation}
\mathbb{E}_{\bar{\mathscr{A}}[\psi]}[\mathscr{O}]=\int_{\Phi|_{\partial\mathscr{M}}=\psi}\mathrm{d}\mu_{(\mathfrak{l}_{\mathrm{UV}},\mathfrak{l}_{\mathrm{IR}})}(\Phi)\,e^{iS_{\mathrm{cl}}[\Phi]/\hbar}\,\mathscr{O}[\Phi],
\end{equation}
may well depend on the regularizing scales $\mathfrak{l}_{\mathrm{UV}}$ and $\mathfrak{l}_{\mathrm{IR}}$. Physical predictions of the quantum theory are necessarily independent of this regularization: the regularization does not represent a physical condition but constitutes a mathematical contrivance. As I discuss shortly, obtaining regularization-independent predictions from the quantum theory is part and parcel of the renormalization process. %I do not wish to dwell on the issue of regularization here: aside from these two points, it is not central to the conceptual points that I aim to convey. 
%In subsubsection \ref{QFTLRNDST} I introduce a particular regularization of the path integration in equation \eqref{QFT1state} and consider its consequences for a renormalization group analysis.

What constitute the physical observables of the quantum theory of the field $\Phi$? If the measure\linebreak $\mathrm{d}\mu_{(\mathfrak{l}_{\mathrm{UV}},\mathfrak{l}_{\mathrm{IR}})}(\Phi)$ respects the invariances of the theory $\mathscr{T}_{\mathrm{cl}}$, then in principle any functional $\mathscr{O}[\Phi]$ of the field $\Phi$ that also respects these invariances counts as a physical observable. If the measure $\mathrm{d}\mu_{(\mathfrak{l}_{\mathrm{UV}},\mathfrak{l}_{\mathrm{IR}})}(\Phi)$ does not respect the invariances of the theory $\mathscr{T}_{\mathrm{cl}}$ (and if one cannot restore these invariances upon renormalization), then in principle any functional $\mathscr{O}[\Phi]$ of the field $\Phi$ that respects the remaining subset of these invariances counts as a physical observable. Determining the subset of invariances respected by the measure $\mathrm{d}\mu_{(\mathfrak{l}_{\mathrm{UV}},\mathfrak{l}_{\mathrm{IR}})}(\Phi)$ and identifying invariant functionals $\mathscr{O}[\Phi]$ (whose physical meaning one comprehends) may also be quite nontrivial problems.

How does one define the modes $\phi_{\ell}$ of scale $\ell$? Since the quantum theory is defined in the presence of the fixed spacetime $\mathscr{M}$ of metric tensor $\mathbf{g}$, one references all scales to those defined by the metric tensor $\mathbf{g}$. One moreover assumes that these metrical scales are measured in units of a chosen standard of length, in this case defined independently of the dynamics of the field $\Phi$. Classically, the field $\Phi$ satisfies the equation of motion $\mathscr{D}\Phi=0$ for the differential operator $\mathscr{D}$, constructed from the metric tensor $\mathbf{g}$, that results from variation of the action $S_{\mathrm{cl}}[\Phi]$. For a free scalar field $\Phi$ of mass $m$, for instance, the equation of motion is the Klein-Gordon equation, $(\nabla^{2}+m^{2})\Phi=0$, where $\nabla^{2}$ is the Laplacian operator of the metric tensor $\mathbf{g}$. One defines the modes $\phi_{\ell}$ of the field $\Phi$ as the eigenfunctions of the differential operator $\mathscr{D}$:
\begin{equation}
\mathscr{D}\phi_{\ell}=\upsilon_{\ell}\phi_{\ell}.
\end{equation}
The eigenvalue $\upsilon_{\ell}$ determines the scale $\ell$ of the mode $\phi_{\ell}$. Again for a free scalar field $\Phi$ of mass $m$, now propagating on Minkowski spacetime, for instance, each eigenvalue has the form $p^{2}+m^{2}$, with $p$ the magnitude of the $d$-momentum vector, which determines the scale $\ell=1/\sqrt{p^{2}+m^{2}}$. 

I am now prepared to define the regularized Lagrangian $\bar{\mathscr{L}}_{\bar{\mathscr{T}}}[\Phi]$ %, which depends on the regulators $\mathfrak{l}_{\mathrm{UV}}$ and $\mathfrak{l}_{\mathrm{IR}}$, 
effective on the interval of scales $(\ell_{\mathrm{UV}},\ell_{\mathrm{IR}})$. The regularized quantum effective action $\bar{S}_{\bar{\mathscr{T}}}[\Phi]$---the spacetime integral of the Lagrangian $\bar{\mathscr{L}}_{\bar{\mathscr{T}}}[\Phi]$---is defined through the relation
\begin{equation}\label{QFT1pathintegral}
e^{i\bar{S}_{\bar{\mathscr{T}}}[\Phi]/\hbar}=\int_{\Phi|_{\partial\mathscr{M}}=\psi}\mathrm{d}\mu_{(\mathfrak{l}_{\mathrm{UV}},\ell_{\mathrm{UV}})\cup(\ell_{\mathrm{IR}},\mathfrak{l}_{\mathrm{IR}})}(\Phi)\,e^{iS_{\mathrm{cl}}[\Phi]/\hbar}
\end{equation}
for the measure
\begin{equation}\label{QFT1regularizedmeasure}
\mathrm{d}\mu_{(\mathfrak{l}_{\mathrm{UV}},\ell_{\mathrm{UV}})\cup(\ell_{\mathrm{IR}},\mathfrak{l}_{\mathrm{IR}})}(\Phi)=\mathscr{N}\prod_{\ell\in(\mathfrak{l}_{\mathrm{UV}},\ell_{\mathrm{UV}})\cup(\ell_{\mathrm{IR}},\mathfrak{l}_{\mathrm{IR}})}\mathrm{d}\phi_{\ell}.
\end{equation}
The action $\bar{S}_{\bar{\mathscr{T}}}[\Phi]$ is a functional of configurations of the field $\Phi$ involving only those modes $\phi_{\ell}$ of scales $\ell\in(\ell_{\mathrm{UV}},\ell_{\mathrm{IR}})$.  %given by the expectation value $\langle\Phi\rangle_{\mathscr{A}}$ of the field $\Phi$ in the quantum state $\mathscr{A}[\phi]$. 
Note that the action $\bar{S}_{\bar{\mathscr{T}}}[\Phi]$ depends on the quantum state through the boundary condition $\Phi|_{\partial\mathscr{M}}=\psi$. Typically, one considers the ground state---and excited states sufficiently close to the ground state---so that this dependence is tacitly ignored. Certain features of the action $\bar{S}_{\bar{\mathscr{T}}}[\Phi]$ are moreover independent of the quantum state. %The path integral measure $\mathrm{d}\mu_{(\mathfrak{l}_{\mathrm{UV}},\ell_{\mathrm{UV}})\cup(\ell_{\mathrm{IR}},\mathfrak{l}_{\mathrm{IR}})}(\Phi)$ is defined as
%\begin{equation}
%\mathrm{d}\mu_{(\mathfrak{l}_{\mathrm{UV}},\ell_{\mathrm{UV}})\cup(\ell_{\mathrm{IR}},\mathfrak{l}_{\mathrm{IR}})}(\Phi)=\prod_{\ell\in(\mathfrak{l}_{\mathrm{UV}},\ell_{\mathrm{UV}})\cup(\ell_{\mathrm{IR}},\mathfrak{l}_{\mathrm{IR}})}\mathrm{d}\phi_{\ell},
%\end{equation}
%where $\phi_{\ell}$ denotes a mode of the field $\Phi$ of length scale $\ell$. One performs the path integration in equation \eqref{QFT1state} only over those modes $\phi_{\ell}$ for which the scale $\ell\in(\mathfrak{l}_{\mathrm{UV}},\ell_{\mathrm{UV}})\cup(\ell_{\mathrm{IR}},\mathfrak{l}_{\mathrm{IR}})$. If one computes a transition amplitude according to the prescription \eqref{QFT1state} using instead the action $S_{\mathscr{T}}[\Phi]$, then this transition amplitude will contain all of the same information as the transition amplitude computed with $S_{\mathrm{cl}}[\Phi]$. %, thereby obtaining a quantum theory effective on the interval of scales $(\ell_{\mathrm{UV}},\ell_{\mathrm{IR}})$.
One may compute the regularized transition amplitude \eqref{QFT1state2} also from the action $\bar{S}_{\bar{\mathscr{T}}}[\Phi]$:
\begin{equation}\label{QFT1statereg}
\bar{\mathscr{A}}[\psi]=\int_{\Phi|_{\partial\mathscr{M}}=\psi}\mathrm{d}\mu_{(\ell_{\mathrm{UV}},\ell_{\mathrm{IR}})}(\Phi)\,e^{i\bar{S}_{\bar{\mathscr{T}}}[\Phi]/\hbar}.
\end{equation}
The dynamics of the field $\Phi$ on the scales $\ell\in(\mathfrak{l}_{\mathrm{UV}},\ell_{\mathrm{UV}})\cup(\ell_{\mathrm{IR}},\mathfrak{l}_{\mathrm{IR}})$ are imprinted on the action $\bar{S}_{\bar{\mathscr{T}}}[\Phi]$ so that the path integrations in equations \eqref{QFT1state2} and \eqref{QFT1statereg} yield the same result. %as the regularized transition amplitude $\bar{\mathscr{A}}[\psi]$, computed as in equation \eqref{QFT1state2} with the action $S_{\mathrm{cl}}[\Phi]$, % with the measure \eqref{QFT1measure1} and 
%now merely restricted to the interval of scales $(\ell_{\mathrm{UV}},\ell_{\mathrm{IR}})$. 

%Specifically, the expectation values of any physical observable $\mathscr{O}$ computed in the quantum states defined by the transition amplitudes $\bar{\mathscr{A}}[\psi]$ and $\bar{\mathscr{A}}_{\bar{\mathscr{T}}}[\psi]$ agree on the interval of scales $(\ell_{\mathrm{UV}},\ell_{\mathrm{IR}})$. (EDIT)

%one decomposes the field $\Phi$ into modes of definite scale as defined by the action of the Laplacian operator on the fields $\Phi$. 

One may rewrite equation \eqref{QFT1pathintegral} as
\begin{equation}\label{QFT1pathintegral'}
e^{i\bar{S}_{\bar{\mathscr{T}}}[\Phi]/\hbar}=\int_{\Phi|_{\partial\mathscr{M}}=\psi}\mathrm{d}\mu_{(\mathfrak{l}_{\mathrm{UV}},\mathfrak{l}_{\mathrm{IR}})}(\Phi)\,\Omega_{(\ell_{\mathrm{UV}},\ell_{\mathrm{IR}})}(\Phi)\,e^{iS_{\mathrm{cl}}[\Phi]/\hbar}
\end{equation}
for an integral kernel $\Omega_{(\ell_{\mathrm{UV}},\ell_{\mathrm{IR}})}(\Phi)$ defined such that
\begin{equation}
\mathrm{d}\mu_{(\mathfrak{l}_{\mathrm{UV}},\ell_{\mathrm{UV}})\cup(\ell_{\mathrm{IR}},\mathfrak{l}_{\mathrm{IR}})}(\Phi)=\mathrm{d}\mu_{(\mathfrak{l}_{\mathrm{UV}},\mathfrak{l}_{\mathrm{IR}})}(\Phi)\,\Omega_{(\ell_{\mathrm{UV}},\ell_{\mathrm{IR}})}(\Phi).
\end{equation}
The integral kernel $\Omega_{(\ell_{\mathrm{UV}},\ell_{\mathrm{IR}})}(\Phi)$ serves to prevent modes $\phi_{\ell}$ of the field $\Phi$ on scales $\ell\in(\ell_{\mathrm{UV}},\ell_{\mathrm{IR}})$ from being integrated out---or, equivalently, to select modes $\phi_{\ell}$ of the field $\Phi$ on scales $\ell\in(\mathfrak{l}_{\mathrm{UV}},\ell_{\mathrm{UV}})\cup(\ell_{\mathrm{IR}},\mathfrak{l}_{\mathrm{IR}})$ for being integrated out---in the path integration of equation \eqref{QFT1pathintegral'}. The integral kernel $\Omega_{(\ell_{\mathrm{UV}},\ell_{\mathrm{IR}})}(\Phi)$ must satisfy the condition
\begin{equation}\label{QFT1condition}
\int_{\Phi|_{\partial\mathscr{M}}=\psi}\mathrm{d}\mu_{(\mathfrak{l}_{\mathrm{UV}},\mathfrak{l}_{\mathrm{IR}})}(\Phi)\,\Omega_{(\ell_{\mathrm{UV}},\ell_{\mathrm{IR}})}(\Phi)=1
\end{equation}
since one may compute the transition amplitude $\bar{\mathscr{A}}[\psi]$ either from the action $S_{\mathrm{cl}}[\Phi]$ as in equation \eqref{QFT1state2} or from the action $\bar{S}_{\bar{\mathscr{T}}}[\Phi]$ as in equation \eqref{QFT1statereg}.
%(COMMENT on properties of kernel, preservation of path integral)

\paragraph{Renormalization}

The Lagrangian $\bar{\mathscr{L}}_{\bar{\mathscr{T}}}[\Phi]$ still depends on the regularizing scales $\mathfrak{l}_{\mathrm{UV}}$ and $\mathfrak{l}_{\mathrm{IR}}$. There are various methods to effect the removal of this dependence from the Lagrangian $\bar{\mathscr{L}}_{\bar{\mathscr{T}}}[\Phi]$---that is, to renormalize the Lagrangian $\bar{\mathscr{L}}_{\bar{\mathscr{T}}}[\Phi]$---thereby obtaining the Lagrangian $\mathscr{L}_{\mathscr{T}}[\Phi]$. (I explain below why this process is one of renormalization.) The addition of counterterms is a particularly transparent method. For each term $\bar{c}_{j}f_{j}[\Phi]$ in the Lagrangian $\bar{\mathscr{L}}_{\bar{\mathscr{T}}}[\Phi]$, %with $\bar{c}_{j}=\bar{c}_{j}(\mathfrak{l}_{\mathrm{UV}},\mathfrak{l}_{\mathrm{IR}})$,
one adds a counterterm $\hat{c}_{j}f_{j}[\Phi]$ such that $c_{j}=\bar{c}_{j}+\hat{c}_{j}$ equals the experimentally measured value of $c_{j}$ on the interval of scales $(\ell_{\mathrm{UV}},\ell_{\mathrm{IR}})$. The couplings $\bar{c}_{j}$, which depend on the regulators $\mathfrak{l}_{\mathrm{UV}}$ and $\mathfrak{l}_{\mathrm{IR}}$, serve to parametrize the contributions %The dependence on the regulators $\mathfrak{l}_{\mathrm{UV}}$ and $\mathfrak{l}_{\mathrm{IR}}$ parametrizes the contributions 
to physics on the interval of scales $(\ell_{\mathrm{UV}},\ell_{\mathrm{IR}})$ from physics on the interval of scales $(0,\mathfrak{l}_{\mathrm{UV}})\cup(\mathfrak{l}_{\mathrm{IR}},\infty)$; the couplings $\hat{c}_{j}$, which characterize the counterterms, serve to match the predictions deriving from the Lagrangian $\bar{\mathscr{L}}_{\bar{\mathscr{T}}}[\Phi]$ to the results coming from experiment. As long as the number of these matchings is finite, the theory $\mathscr{T}$ is predictive. Once one has performed this matching on the interval of scales $(\ell_{\mathrm{UV}},\ell_{\mathrm{IR}})$, one may employ the Lagrangian $\mathscr{L}_{\mathscr{T}}[\Phi]$ to derive regularization-independent predictions concerning physical observables (not already measured for the purpose of matching) on scales $\ell\in(\ell_{\mathrm{UV}},\ell_{\mathrm{IR}})$. As I explain shortly, one may then apply the renormalization group to derive regularization-independent predictions concerning physical observables on all other scales.

%As I explained in subsubsection \ref{conceptualRG}, 
One does not expect to be able to predict the empirical values of the couplings $c_{j}$ on the interval of scales $(\ell_{\mathrm{UV}},\ell_{\mathrm{IR}})$; %; rather, one requires experimental input to set the values of the couplings characterizing relevant scalar functionals.\footnote{In the absence of a fixed point theory $\mathscr{T}_{*}$, one requires experimental input to set the values of all of the couplings $c_{j}$ at the scale $\ell$.} 
rather, one expects to require experimental input to specify the quantum theory under consideration. The basis of these expectations is plain: before performing any measurements, one does not know which renormalization group trajectory one's experiment probes. A measurement of the couplings on the interval of scales $(\ell_{\mathrm{UV}},\ell_{\mathrm{IR}})$ serves to specify this renormalization group trajectory.
%values of the couplings of relevant scalar functionals at some scale $\ell$.
%As I explained in subsection \ref{conceptualRG}, a renormalization group trajectory specifies a quantum theory of the field $\Phi$. 
These expectations persist even if one (somehow) knows that the renormalization group trajectory being experimentally probed contains a fixed point. Although one can compute the values $c_{j}^{*}$ of the couplings at this fixed point, % from the form of the renormalization group transformation, 
one again does not know which of the renormalization group trajectories spanning the critical surface one's experiment probes.
%In the presence of a fixed point theory $\mathscr{T}_{*}$, the renormalization group transformations determine the values of the couplings of irrelevant scalar functionals starting from their vanishing values at the fixed point. In the absence of a fixed point theory $\mathscr{T}_{*}$, one requires experimental input to set the values of all of the couplings $c_{j}$ at some scale $\ell$ since one cannot classify scalar functionals as relevant or irrelevant. As long as the number of couplings to be determined empirically is finite, the theory $\mathscr{T}$ is predictive in the sense that one can predict the values of the couplings at other scales using the renormalization group.
%Although one can compute the values $c_{j}^{*}$ of the couplings at a fixed point from the quantum theory and the form of the renormalization group transformation, (in a typical experimental setting) one must measure the values of the couplings at some scale $\ell$ and then use the renormalization group transformation to flow these values to other scales. A typical renormalization group trajectory is not tuned to criticality, so, although it might pass near a fixed point, it does not intersect the fixed point. A measurement of the couplings serves to pick out the renormalization group trajectory being experimentally probed. 
Since the relevant scalar functionals yield the dominant contributions to the dynamics of the field $\Phi$ on the interval of scales controlled by this fixed point, one typically counts the number of couplings to be determined experimentally as the number of relevant scalar functionals. If one can probe subdominant contributions to the dynamics, still on the interval of scales controlled by this fixed point, then one must also fix experimentally the couplings characterizing irrelevant scalar functionals. 

%are considered to be parameters that must be fixed by experiment (at least on the interval of scales controlled by the fixed point). 

%Renormalization group transformation then flow these empirically determined values to other scales. 

Having now determined the Lagrangian $\mathscr{L}_{\mathscr{T}}[\Phi]$, one can apply renormalization group transformations to obtain the %If one had the renormalization group flows of the couplings $c_{j}$, then one could obtain the 
Lagrangians $\mathscr{L}_{\mathscr{T}'}[\Phi]$ effective on other intervals of scales. Knowledge of the Lagrangians $\mathscr{L}_{\mathscr{T}'}[\Phi]$ is tantamount to knowledge of the renormalization group flows %  Suppose that one wishes to compute the renormalization group flows 
of the couplings $c_{j}$ parametrizing the space $\mathfrak{T}$ of theories $\mathscr{T}$ of the field $\Phi$. %of the Lagrangian $\mathcal{L}_{\bar{\mathscr{T}}}[\Phi]$ 
Consider in particular a renormalization group transformation that incrementally increases the ultraviolet scale $\ell_{\mathrm{UV}}$ and leaves fixed the infrared scale $\ell_{\mathrm{IR}}$. Implementing this renormalization group transformation is conceptually straightforward: integrate out all modes $\phi_{\ell}$ of the field $\Phi$ on the interval of scales $(\ell_{\mathrm{UV}},\ell_{\mathrm{UV}}+\delta\ell)$. The resulting Lagrangian $\mathscr{L}_{\mathscr{T}'}[\Phi]$ is formally defined through the relation
\begin{equation}\label{QFT1RGtrans}
e^{iS_{\mathscr{T}'}[\Phi]/\hbar}=\int_{\Phi|_{\partial\mathscr{M}}=\phi}\mathrm{d}\mu_{(\ell_{\mathrm{UV}},\ell_{\mathrm{UV}}+\delta\ell)}(\Phi)\,e^{iS_{\mathscr{T}}[\Phi]/\hbar}.
\end{equation}
This renormalization group transformation automatically maps the theory $\mathscr{T}$ into the theory $\mathscr{T}'$ describing the same physics: one has explicitly identified the modes within the interval of scales $(\ell_{\mathrm{UV}},\ell_{\mathrm{UV}}+\delta\ell)$ and integrated them out, leaving fixed the modes within the interval of scales $(\ell_{\mathrm{UV}}+\delta\ell,\ell_{\mathrm{IR}})$. One may alternatively formulate  the renormalization group transformation of expression \eqref{QFT1RGtrans} in terms of an integral kernel $\Omega(\Phi)$: to implement this renormalization group transformation, one inserts the integral kernel $\Omega_{(\ell_{\mathrm{UV}}+\delta\ell,\ell_{\mathrm{IR}})}(\Phi)$ into the path integral. One may conceive of the integral kernel $\Omega_{(\ell_{\mathrm{UV}}+\delta\ell,\ell_{\mathrm{IR}})}(\Phi)$ as encoding the coarse graining operation executed on the field $\Phi$. The condition \eqref{QFT1condition} dictates that a coarse graining operation does not change the value of the transition amplitude $\mathscr{A}[\psi]$. One extracts the renormalization group flows of the couplings $c_{j}$ by iterating the renormalization group transformation.

One now achieves an understanding of the origin of the formal divergence of the path integral, %in equation \eqref{QFT1state}, 
the role of the regularizing scales $\mathfrak{l}_{\mathrm{UV}}$ and $\mathfrak{l}_{\mathrm{IR}}$, and the removal of the regularization by a process of renormalization. %is thus laid bare: 
In the path integration of equation \eqref{QFT1state}, one simply assumes that the quantum theory so defined %encoded in the action $S_{\mathrm{cl}}[\Phi]$ 
applies on all scales. This represents a significant extrapolation of one's knowledge %: in actuality one only knows that this physics applies to an interval of scales bounded from above and from below. One is thus extrapolating %it is merely a result of one's extrapolation 
of physics on scales about which one is not ignorant to physics on scales about which one is ignorant. There is no guarantee that this extrapolation is valid, and the divergence of the path integral indicates that it is not. The regulators $\mathfrak{l}_{\mathrm{UV}}$ and $\mathfrak{l}_{\mathrm{IR}}$ curtail this extrapolation so that divergences do not arise. The specific values of the regulators $\mathfrak{l}_{\mathrm{UV}}$ and $\mathfrak{l}_{\mathrm{IR}}$ are largely arbitrary: one needs to cut off the extrapolation in the ultraviolet at some scale $\ell_{\mathrm{UV}}$ and in the infrared at some scale $\ell_{\mathrm{IR}}$. There is thus no conceptual difference between the regulator $\mathfrak{l}_{\mathrm{UV}}$ and the scale $\ell_{\mathrm{UV}}$ and between the regulator $\mathfrak{l}_{\mathrm{IR}}$ and the scale $\ell_{\mathrm{IR}}$. One could have taken $\mathfrak{l}_{\mathrm{UV}}=\ell_{\mathrm{UV}}$ and $\mathfrak{l}_{\mathrm{IR}}=\ell_{\mathrm{IR}}$ from the start. 
%Since one's extrapolation is clearly incorrect if divergences result, one is truly ignorant of physics on some scales. 
Lacking a knowledge of physics on scales $\ell\in(0,\mathfrak{l}_{\mathrm{UV}})\cup(\mathfrak{l}_{\mathrm{IR}},\infty)$, one can nevertheless parametrize one's ignorance by matching the predictions of the quantum theory to the outcomes of experimental measurements at the scales $\mathfrak{l}_{\mathrm{UV}}$ and $\mathfrak{l}_{\mathrm{IR}}$. (I assume that the quantum theory accurately describes physics on scales $\ell\in(\mathfrak{l}_{\mathrm{UV}},\mathfrak{l}_{\mathrm{IR}})$.) %since one could perform the matching for any values. 
One can now employ the renormalization group to extrapolate the predictions of the renormalized quantum theory to scales $\ell\in(0,\mathfrak{l}_{\mathrm{UV}})\cup(\mathfrak{l}_{\mathrm{IR}},\infty)$, and one can check these predictions against the outcomes of experimental measurements. %Renormalization group transformations treat the regulators $\mathfrak{l}_{\mathrm{UV}}$ and $\mathfrak{l}_{\mathrm{IR}}$ as they do any other pair of ultraviolet and infrared scales. 
If one finds an ultraviolet fixed point (by iterating the scale $\ell_{\mathrm{UV}}$ to zero), then the path integration in equation \eqref{QFT1state} was not in fact ultraviolet-divergent, and one did not require the regulator $\mathfrak{l}_{\mathrm{UV}}$. If one finds an infrared fixed point (by iterating the scale $\ell_{\mathrm{IR}}$ to infinity), then the path integration in equation \eqref{QFT1state} was not in fact infrared-divergent, and one did not require the regulator $\mathfrak{l}_{\mathrm{IR}}$. %(completely correct? what about shift from counterterms?)

\subsubsection{Quantum field theory on a lattice-regularized nondynamical spacetime}\label{QFTLRNDST}

\paragraph{Definition}

As I discussed in subsubsection \ref{QFTCNDST}, one must regularize the path integration in equation \eqref{QFT1state} to render it well-defined. If one wishes to continue to work in the continuum, then usually one can only carry out this regularization in a perturbative setting close to a fixed point of the quantum theory.\footnote{Functional renormalization group techniques represent an exception, but in this setting one must restrict the Lagrangians $\mathscr{L}_{\mathscr{T}}[\Phi]$ to those obtained from a truncation $\mathfrak{t}$ of the space $\mathfrak{T}$ chosen beforehand \cite{MN&MR}.} To explore nonperturbatively the dynamics of the field $\Phi$, one often employs numerical methods to study a lattice regularization of the path integration in equation \eqref{QFT1state}. Although one requires the lattice regularization to obtain a well-defined quantum theory, one's interests actually lie with the limit of this theory---if it exists---in which one removes the regularization in such a manner that physical observables remain finite. This limit is a continuum limit if the requisite fixed point theories exist or a continuum description if the requisite fixed point theories do not exist.\footnote{The former expression---continuum limit---is standard while the latter expression---continuum description---is not. I introduce the latter as a convenient converse to the former.} %Even if this limit does not exist, one may still be able to obtain predictions concerning the nonperturbative dynamics of the field $\Phi$ within a continuous interval of scales.  (distinguish continuum limit from continuum description) %continuum description, an effective field theory. 

Consider accordingly the quantum theory of the field $\Phi$ propagating on a fixed lattice-regularized spacetime $\mathcal{M}$. The choice of a lattice regularization and the assignment of a lattice spacing $a$ determine the metrical structure of the spacetime $\mathcal{M}$. For instance, one might take $\mathcal{M}$ to be a regular quadrangulation of Minkowksi spacetime. The lattice spacing's %serves as the chosen standard of length. Its 
value, however, is \emph{a priori} arbitrary in that there exists no \emph{a priori} connection between the unit of length defined by the lattice spacing and a physical unit of length such as the meter. %One is nevertheless licensed in using the lattice spacing as a reference scale because it characterizes a fixed background structure in the formulation of the theory. 
How does the analysis of subsubsection \ref{QFTCNDST} carry over to this setting?

Starting from a classical theory $\mathscr{T}_{\mathrm{cl}}$ of the field $\Phi$ specified by the Lagrangian $\mathscr{L}_{\mathrm{cl}}[\Phi]$, one first constructs a corresponding discrete Lagrangian $\mathcal{L}_{\mathrm{cl}}[\Phi]$ suited to the chosen lattice regularization. This construction is generally not unique since many discrete Lagrangians $\mathcal{L}_{\mathrm{cl}}[\Phi]$ coincide with the continuous Lagrangian $\mathscr{L}_{\mathrm{cl}}[\Phi]$ in the limit of vanishing lattice spacing. This construction also does not generally respect the invariances of the theory $\mathscr{T}_{\mathrm{cl}}$ since these invariances may not have a discrete realization. Only after studying the continuum description of the quantum theory defined by equation \eqref{QFT2state} below can one determine whether or not one has employed an appropriate discrete Lagrangian. At the very least one judges the appropriateness of a discrete Lagrangian by whether or not the classical limit of the continuum description coincides with the classical theory $\mathscr{T}_{\mathrm{cl}}$. Fortunately, a continuum limit, though not a generic continuum description, is insensitive to many details of the discrete Lagrangian---it exhibits so-called universality. Specifically, certain details of the Lagrangian $\mathcal{L}_{\mathrm{cl}}[\Phi]$ give rise to irrelevant scalar functionals (relative to a fixed point theory $\mathscr{T}_{*}$) in a Lagrangian $\mathscr{L}_{\mathscr{T}}[\Phi]$ providing part of the continuum description. 
%Although the lattice regularization renders the theory $\mathscr{T}$ well-defined, a whole host of new challenges arise. 
All dimensionful quantities in the Lagrangian $\mathcal{L}_{\mathrm{cl}}[\Phi]$ appear by construction in units of the lattice spacing $a$. For a scalar field $\Phi$ of mass $m$, for instance, the field $\Phi$ appears as $\tilde{\Phi}a^{-1}$ and the mass $m$ appears as $\tilde{m}a^{-1}$. The Lagrangian $\mathcal{L}_{\mathrm{cl}}[\Phi]$ itself carries units of $a^{-(d+1)}$. The discrete classical action $\mathcal{S}_{\mathrm{cl}}[\Phi]$---the spacetime sum of the Lagrangian $\mathcal{L}_{\mathrm{cl}}[\Phi]$---is dimensionless (in units for which $\hbar=1$). One thus expresses the action $\mathcal{S}_{\mathrm{cl}}[\Phi]$ in terms of dimensionless quantities. Again for a scalar field $\Phi$ of mass $m$, for instance, these dimensionless quantities are precisely $\tilde{\Phi}$ and $\tilde{m}$. %For a free scalar field $\Phi$ of mass $m$, the analogue of $\Phi$ is $\tilde{\Phi}$, and the analogue of $m$ is $\tilde{m}$. 

One now computes a transition amplitude $\mathcal{A}[\tilde{\psi}]$ in the quantum theory as the path sum
\begin{equation}\label{QFT2state}
\mathcal{A}[\tilde{\psi}]=\sum_{\substack{\tilde{\Phi} \\ \tilde{\Phi}|_{\partial\mathcal{M}}=\tilde{\psi}}}\mu_{(1,\tilde{L})}(\tilde{\Phi})\,e^{i\mathcal{S}_{\mathrm{cl}}[\tilde{\Phi}]/\hbar}.
\end{equation}
The configuration $\tilde{\psi}$ of the field $\tilde{\Phi}$ on the boundary $\partial\mathcal{M}$ of the spacetime $\mathcal{M}$ specifies the transition amplitude $\mathcal{A}[\tilde{\psi}]$. The expression \eqref{QFT2state} is an instruction to sum over all configurations of the field $\tilde{\Phi}$ satisfying the boundary condition $\tilde{\Phi}|_{\partial\mathcal{M}}=\tilde{\psi}$, weighting each configuration by the product of the measure $\mu_{(1,\tilde{L})}(\tilde{\Phi})$ and the exponential $e^{i\mathcal{S}_{\mathrm{cl}}[\tilde{\Phi}]/\hbar}$. The measure $\mu_{(1,\tilde{L})}(\tilde{\Phi})$ indicates the presence of the lattice spacing $a$---unity in units of the lattice spacing---serving as an ultraviolet regulator $\mathfrak{l}_{\mathrm{UV}}$ and the lattice extent $L$---$\tilde{L}$ in units of the lattice spacing---serving as an infrared regulator $\mathfrak{l}_{\mathrm{IR}}$. %One often choses the explicit form of the measure $\mu_{(1,\tilde{L})}(\tilde{\Phi})$ to eliminate residual redundancies of description not already eliminated by the lattice regularization. 
There is again some freedom in the definition of the measure $\mu_{(1,\tilde{L})}(\tilde{\Phi})$, and one's choice of measure may well affect the resulting continuum description. At the very least one designs the measure to eliminate residual redundancies of description not already eliminated by the lattice regularization.  

In analogy to equation \eqref{QFT1measure1}, one may in principle define the measure $\mu_{(1,\tilde{L})}(\tilde{\Phi})$ in terms of the modes $\tilde{\phi}_{\tilde{\ell}}$
of the field $\tilde{\Phi}$ of length scale $\tilde{\ell}$. How does one define the modes $\tilde{\phi}_{\tilde{\ell}}$ (in units of the lattice spacing $a$)? %in units of the lattice spacing $a$? 
Since the theory is still defined in the presence of a fixed spacetime $\mathcal{M}$, one still references all scales to those defined by its metrical structure. Classically, the field $\tilde{\Phi}$ satisfies the equation of motion $\mathcal{D}\tilde{\Phi}=0$ for the difference operator $\mathcal{D}$, constructed from the spacetime $\mathcal{M}$, that results from variation of the action $\mathcal{S}_{\mathrm{cl}}[\tilde{\Phi}]$. One defines the modes $\tilde{\phi}_{\tilde{\ell}}$ of the field $\tilde{\Phi}$ as the eigenvectors of the difference operator $\mathcal{D}$:
\begin{equation}\label{QFT2modes}
\mathcal{D}\tilde{\phi}_{\tilde{\ell}}=\tilde{\upsilon}_{\tilde{\ell}}\tilde{\phi}_{\tilde{\ell}}.
\end{equation}
The eigenvalue $\tilde{\upsilon}_{\tilde{\ell}}$ specifies the scale $\tilde{\ell}$ of the mode $\tilde{\phi}_{\tilde{\ell}}$, which is restricted to the interval $(1,\tilde{L})$. %$\{a,2a,\ldots,(\tilde{L}-1)a,\tilde{L}a\}$. %I introduce the dimensionless length scale $\tilde{l}$ defined by $\ell=\tilde{\ell}a$. 
For a free scalar field $\tilde{\Phi}$ of mass $\tilde{m}$ propagating on a regular quadrangulation of Minkowski spacetime, for instance, each eigenvalue has the form $\tilde{p}^{2}+\tilde{m}^{2}$ for $\tilde{p}^{2}=4\sum_{j=1}^{d}\sin^{2}{\left(\frac{\tilde{p}_{j}}{2}\right)}$ with $\tilde{p}_{j}\in\left\{0,\frac{2\pi}{\tilde{L}},\ldots,\frac{2\pi(\tilde{L}-1)}{\tilde{L}}\right\}$, which determines the scale $\tilde{\ell}=1/\sqrt{\tilde{p}^{2}+\tilde{m}^{2}}$ \cite{HJR}. 

%Introducing such a lattice regularization transforms the relation \eqref{QFT1pathintegral} into
I am now prepared to define the Lagrangian $\mathcal{L}_{\bar{\mathscr{T}}}[\tilde{\Phi}]$ effective on the interval of scales $(\tilde{\ell}_{\mathrm{UV}},\tilde{\ell}_{\mathrm{IR}})$ for $1\leq\tilde{\ell}_{\mathrm{UV}}\leq\tilde{\ell}_{\mathrm{IR}}\leq\tilde{L}$. The quantum effective action $\mathcal{S}_{\bar{\mathscr{T}}}[\tilde{\Phi}]$---the spacetime sum of the Lagrangian $\mathcal{L}_{\bar{\mathscr{T}}}[\tilde{\Phi}]$---is defined through the relation
\begin{equation}\label{QFT1pathsum}
e^{i\mathcal{S}_{\bar{\mathscr{T}}}[\tilde{\Phi}]/\hbar}=\sum_{\substack{\tilde{\Phi} \\ \tilde{\Phi}|_{\partial\mathcal{M}}=\tilde{\psi}}}\mu_{(1,\tilde{\ell}_{\mathrm{UV}})\cup(\tilde{\ell}_{\mathrm{IR}},\tilde{L})}(\tilde{\Phi})\,e^{i\mathcal{S}_{\mathrm{cl}}[\tilde{\Phi}]/\hbar}.
\end{equation}
%for the measure
%\begin{equation}
%\mu_{(1,\tilde{\ell}_{\mathrm{UV}})\cup(\tilde{\ell}_{\mathrm{IR}},\tilde{L})}(\tilde{\Phi})=\prod_{\tilde{\ell}\in(1,\tilde{\ell}_{\mathrm{UV}})\cup(\tilde{\ell}_{\mathrm{IR}},\tilde{L})}\tilde{\phi}_{\tilde{\ell}}.
%\end{equation}
One may compute the transition amplitude \eqref{QFT2state} also from the action $\mathcal{S}_{\bar{\mathscr{T}}}[\tilde{\Phi}]$:
\begin{equation}\label{QFT2state2}
\mathcal{A}[\tilde{\psi}]=\sum_{\substack{\tilde{\Phi} \\ \tilde{\Phi}|_{\partial\mathcal{M}}=\tilde{\psi}}}\mu_{(\tilde{\ell}_{\mathrm{UV}},\tilde{\ell}_{\mathrm{IR}})}(\tilde{\Phi})\,e^{i\mathcal{S}_{\bar{\mathscr{T}}}[\tilde{\Phi}]/\hbar}.
\end{equation}
The dynamics of the field $\tilde{\Phi}$ on the scales $\ell\in(1,\tilde{\ell}_{\mathrm{UV}})\cup(\tilde{\ell}_{\mathrm{IR}},\tilde{L})$ are imprinted on the action $\mathcal{S}_{\bar{\mathscr{T}}}[\tilde{\Phi}]$ so that the path summations in equations \eqref{QFT2state} and \eqref{QFT2state2} yield the same result. % the same information as the transition amplitude $\mathcal{A}[\tilde{\psi}]$, computed in equation \eqref{QFT2state} with the action $\mathcal{S}_{\mathrm{cl}}[\tilde{\Phi}]$, now merely restricted to the interval of scales $(\tilde{\ell}_{\mathrm{UV}},\tilde{\ell}_{\mathrm{IR}})$. 
The path summations in equations \eqref{QFT2state}, \eqref{QFT1pathsum}, and \eqref{QFT2state2}, as well as the multidimensional diagonalization required to solve the characteristic equation \eqref{QFT2modes}, often prove analytically intractable, so one resorts to numerical methods. Markov chain Monte Carlo simulations constitute a standard technique, indeed, essentially the standard technique. One begins by performing a Wick rotation from the Lorentzian to the Euclidean sector, transforming the complex weights $\mu(\tilde{\Phi})\,e^{i\mathcal{S}[\tilde{\Phi}]/\hbar}$ of the path sum into the real weights $\mu^{(\mathrm{E})}(\tilde{\Phi})\,e^{-\mathcal{S}^{(\mathrm{E})}[\tilde{\Phi}]/\hbar}$ of a partition function $\mathcal{Z}[\tilde{\psi}]$. This Wick rotation is only well-defined for a suitable fixed spacetime $\mathcal{M}$, a suitable measure $\mu(\tilde{\Phi})$, and a suitable action $\mathcal{S}_{\mathrm{cl}}[\tilde{\Phi}]$. One thus obtains a statistical mechanical model defined by the partition function $\mathcal{Z}[\tilde{\psi}]$ to study as such. 

A Markov chain Monte Carlo simulation then produces an ensemble of configurations of the field $\tilde{\Phi}$ representative of those contributing to the partition function $\mathcal{Z}[\tilde{\psi}]$. A particular ensemble corresponds to a particular choice of the bare couplings $\tilde{c}_{j}$ of the Lagrangian $\mathcal{L}_{\mathrm{cl}}^{(\mathrm{E})}[\tilde{\Phi}]$ as well as the lattice extent $\tilde{L}$. By performing numerical measurements of discrete observables $\mathcal{O}$ on this ensemble, one gleans information about the partition function $\mathcal{Z}[\tilde{\psi}]$ and the Lagrangians $\mathcal{L}_{\bar{\mathscr{T}}}^{(\mathrm{E})}[\tilde{\Phi}]$. Specifically, one has access to the averages 
\begin{equation}\label{ensembleaverage}
\langle\mathcal{O}\rangle_{\mathcal{A}[\tilde{\psi}]}=\sum_{j=1}^{N(\tilde{\Phi})}\mathcal{O}_{j}
\end{equation}
of the observables $\mathcal{O}$ over an ensemble of $N(\tilde{\Phi})$ configurations of the field $\tilde{\Phi}$. The ensemble average \eqref{ensembleaverage} approximates the expectation value 
\begin{equation}
\mathbb{E}_{\mathcal{A}[\tilde{\psi}]}[\mathcal{O}]=\sum_{\substack{\tilde{\Phi} \\ \tilde{\Phi}|_{\partial\mathcal{M}}=\tilde{\psi}}}\mu_{(1,\tilde{L})}^{(\mathrm{E})}(\tilde{\Phi})\,e^{-\mathcal{S}_{\mathrm{cl}}^{(\mathrm{E})}[\tilde{\Phi}]/\hbar}\mathcal{O}[\tilde{\Phi}]
\end{equation}
as guaranteed by the Metropolis (or related) algorithm employed in Markov chain Monte Carlo simulations. Although the expectation values of a complete set of observables $\mathcal{O}$ completely determine the partition function $\mathcal{Z}[\tilde{\psi}]$, in practice one cannot obtain such complete information. Moreover, even if one manages to reconstruct the partition function $\mathcal{Z}[\tilde{\psi}]$, the implications of the physics contained within the partition function $\mathcal{Z}[\tilde{\psi}]$ for the physics contained within the path sum $\mathcal{A}[\tilde{\psi}]$ may be far from clear unless an Osterwalder-Schrader-type theorem holds \cite{KO&RS1,KO&RS2}.

%Since the partition function $\mathcal{Z}[\tilde{\phi}]$ defines a statistical mechanical model, one studies it in this context. 

%One aims to determine the continuum limit of the theory $\mathscr{T}$ defined by the lattice regularization, obtained by appropriating letting the number of lattice constituents increase without bound and the lattice spacing decrease to zero. 

%it adds the complications arising from a finite lattice spacing and a finite lattice extent. One is then interested in the continuum limit of the lattice theory---if it exists---in which the lattice spacing vanishes and the lattice grows in proportion. One must infer properties of the continuum limit by performing finite size scaling analyses. One does not know what the continuum theory is, so in analyzing data one must make an assumption about what theory to use and then revise that assumption depending on the outcome of statistical analyses of data. What one considers to be a physical observable depends on the theory that one chooses to use. 

%Since the lattice in this case is still a fixed background, one is licensed in assuming that a coarse graining step increases the lattice spacing. Moreover, if one is considering a theory the continuum limit of which likely describes some actual phenomenon, then one can eventually bring the lattice spacing into contact with physically defined scales by making contact with known phenomena. 

\paragraph{Renormalization}

As I emphasized above, one in fact wishes to glean information about the Lagrangians $\mathscr{L}_{\mathscr{T}}^{(\mathrm{E})}[\Phi]$---and eventually their Lorentzian counterparts---providing the continuum description of the Lagrangians $\mathcal{L}_{\bar{\mathscr{T}}}^{(\mathrm{E})}[\tilde{\Phi}]$. % interests actually lie with the continuum description not with the regulated theories. 
Since the Lagrangians $\mathcal{L}_{\bar{\mathscr{T}}}^{(\mathrm{E})}[\tilde{\Phi}]$ are defined in the presence of a regularization, one should be able to access the Lagrangians $\mathscr{L}_{\mathscr{T}}^{(\mathrm{E})}[\Phi]$ through a renormalization process. With the lattice spacing $a$ serving as an ultraviolet regulator and the lattice extent $L$ serving as an infrared regulator, one thus considers a limit in which the lattice spacing $a$ decreases to zero and the lattice extent $L$ increases without bound while physical quantities remain finite. %If the associated renormalization group flow approaches a fixed point in the ultraviolet, then one can take the limit of vanishing lattice spacing. If the associated renormalization group flow approaches a fixed point in the infrared, then one can take the limit of infinite lattice extent. If the associated renormalization group flow does not contain one or both of these fixed points, then one can likely still obtain a continuum description effective on a nontrivial interval of scales. 
Given only an ensemble of configurations of the field $\tilde{\Phi}$ generated by Markov chain Monte Carlo simulations, %Suppose now that one wishes to compute the renormalization group flows of the couplings $c_{j}$ of the continuum limit for a renormalization group transformation that incrementally increases the ultraviolet scale $\ell_{\mathrm{UV}}$ and leaves fixed the infrared scale $\ell_{\mathrm{IR}}$. Implementing this renormalization group transformation is no longer so straightforward. If one had analytic control over the mode decomposition \eqref{QFT2modes} and the transition amplitude $\mathcal{A}[\tilde{\psi}]$, then one could straightforwardly sum out modes on an interval of scales $(\tilde{\ell}_{\mathrm{UV}},\tilde{\ell}_{\mathrm{UV}}+\delta\tilde{\ell})$ in complete analogy to equation \eqref{QFT1RGtrans}. If one instead has only an ensemble of configurations of the field $\tilde{\Phi}$ generated by Markov chain Monte Carlo simulations, then one can neither readily identify the modes $\tilde{\phi}_{\tilde{\ell}}$ nor compute the partition function $\mathcal{Z}[\tilde{\psi}]$. %One cannot simply integrate out modes of the field $\Phi$ on the scales $(\ell_{\mathrm{UV}},\ell_{\mathrm{UV}}+\delta\ell)$ because one does not know how to identify these modes. (One can simply integrate out modes of the field by coarse graining, but one needs to make sure that the resulting theory describes the same dynamics.)
one requires a renormalization group scheme of a nature quite different from that discussed in subsubsection \ref{QFTCNDST}. I now explain how such a scheme typically works. 

One first selects a candidate model for the continuum description of the lattice-regularized quantum theory. %(DEFINE how different from theory) %---either a renormalized effective theory or a true continuum limit---
%in terms of which one analyzes and interprets numerical measurements of discrete observables. 
This model consists of a truncation $\mathfrak{t}$ of the space $\mathfrak{T}$ of theories $\mathscr{T}$, %defined by a specific subset of the scalar functionals $f_{j}[\mathbf{g}]$, 
each theory $\mathscr{T}$ specified by a Lagrangian $\mathscr{L}_{\mathscr{T}}^{(\mathrm{E})}[\Phi]$ of the form \eqref{lagrangian} for as yet undetermined values of the couplings $c_{j}$.\footnote{The choice of a truncation has a different status in this setting to that of the functional renormalization group: as I explain shortly, one ascertains the appropriateness of the chosen truncation by its ability to account for numerical data.} As the notation indicates, in terms of the discussion of renormalization of subsubsection \ref{QFTCNDST}, one should consider the Lagrangian $\mathscr{L}_{\mathscr{T}}^{(\mathrm{E})}[\Phi]$ to include already contributions from counterterms. %The couplings $c_{j}$ that one aims to extract are thus the renormalized couplings. 
For a real scalar field $\Phi$ invariant under parity inversion, for instance, one might define a succession of models by the quantum effective actions
\begin{equation}
S_{\mathscr{T}}^{(\mathrm{E})}[\Phi]=\int_{\mathscr{M}}\mathrm{d}^{d+1}x\sqrt{g}\left[c_{0}\Phi^{2}+c_{1}\Phi^{4}+c_{2}\nabla_{a}\Phi\nabla^{a}\Phi+c_{3}\Phi^{6}+c_{4}\Phi^{2}\nabla_{a}\Phi\nabla^{a}\Phi+\cdots\right]%c_{4}\nabla_{a}\nabla_{b}\Phi\nabla^{a}\nabla^{b}\Phi+c_{5}\nabla_{a}\nabla^{a}\Phi\nabla_{b}\nabla_{a}\Phi+c_{6}(\nabla_{a}\Phi\nabla^{a}\Phi)^{2}+\cdots\right]
\end{equation}
expanded to successive powers in the field $\Phi$ and its covariant derivatives. 

Since one in fact wishes to infer the model providing the continuum description of the lattice-regularized quantum theory, why must one select a candidate model beforehand?
%What purpose does such a model serve? 
%Quite simply, one cannot make sense of 
A model establishes the framework within which one analyzes and interprets numerical measurements of discrete observables performed on an ensemble of configurations of the field $\tilde{\Phi}$. For a renormalization group analysis---and, indeed, almost any analysis---this framework has three key facets: first, the physical observables $\mathscr{O}$ of the model, which one wishes to deduce from numerical measurements of discrete observables $\mathcal{O}$; second, the couplings of the model whose renormalization group flows one wishes to extract from numerical measurements; and, third, the measurements of physical observables that yield the values of the couplings, which one wishes to implement on an ensemble of configurations of the field $\tilde{\Phi}$. 
%In particular, a model establishes the quantities that constitute its physical observables $\mathscr{O}$ that one wishes to infer from numerical measurements of discrete observables $\mathcal{O}$ %. Moreover, a model establishes the couplings %A model establishes the couplings 
%and the couplings whose renormalization group flows one wishes to extract and the measurements from whose results one can infer the renormalized values of the couplings. 
%These measurements give physical meaning to the couplings. 
Together these facets constitute the physical input required for a renormalization group analysis. Thus, quite simply, one cannot ascribe physical meaning to the results of numerical measurements in the absence of a model. As I discuss below, the physical observables of a model also establish the meaning of its renormalization group trajectories. Given a model for the continuum description, the determination of its physical observables---and how to measure them---may prove a highly nontrivial task. %Without a model one quite simply cannot give physical meaning to these numerical measurements.

How does one select a model without already knowing what the continuum description is? %One can employ either of two methods, which in the end both involve the same analyses. 
Either one starts from the simplest nontrivial model and blindly iterates through successively more complicated models, or one attempts to determine directly the model that accounts for one's numerical measurements. %Typically, one follows the latter route, banking on one's expectations for the continuum limit. 
When implementing either of these strategies, especially the latter, one relies liberally on one's expectations for the continuum description and on one's intuition about the physical meaning of discrete observables. Either way statistical analyses of fits of selected models to one's numerical measurements ultimately determine the model that one employs. The outcomes of these statistical analyses depend on the numerical measurements that one performs, in particular, on which physical observables one probes and on which length scales one probes. The former dependence is crucial: one must probe a sufficient number of physical observables not only to assess a model's viability, but also to distinguish between models. The latter dependence is expected: the continuum description consists of a succession of theories $\mathscr{T}$ effective on incrementally changing intervals of scales. 

The analysis of numerical measurements in terms of the chosen model requires a finite size scaling \emph{Ansatz}, a prescription for how discrete quantities scale into their continuous counterparts in the continuum description. Since one's interests lie with the continuum description, and one's model is of this continuum description, but one's methods work at finite lattice spacing and extent, one requires a means to address one's methods to one's interests. % for studying the continuum limit while working at finite lattice spacing and finite lattice extent. 
This is precisely the purpose of finite size scaling analysis. The selected model dictates a category of finite size scaling \emph{Ans\"atze} consisting of those compatible with the model's physical content. Typically, a finite size scaling \emph{Ansatz} is based on a dimensionful physical observable $\mathscr{O}$ of the chosen model. One obtains the value of the physical observable $\mathscr{O}$ from its discrete analogue $\mathcal{O}$ \emph{via} the relation
\begin{equation}\label{FSSAnsatz}
\mathscr{O}=\lim_{a\rightarrow0}\mathcal{O}a^{q},
\end{equation}
where the power $q$ is the dimension of the physical observable $\mathscr{O}$. (This limit might also involve letting the dimensionless lattice extent $\tilde{L}$ increase without bound.) For instance, if the chosen model is a quantum theory of a scalar field $\Phi$ of mass $m$, then one might take the mass $m$ as the physical observable $\mathscr{O}$. By measuring the $2$-point correlation function of the scalar field $\tilde{\Phi}$, one determines the correlation length $\tilde{\xi}$, related to the mass $m$ as
\begin{equation}\label{correlationlength}
m=\lim_{a\rightarrow0}\frac{1}{\tilde{\xi} a}.
\end{equation}
For the physical observable $\mathscr{O}$ to remain finite in the limit of vanishing lattice spacing, the discrete observable $\mathcal{O}$ must diverge appropriately. Discrete observables of the lattice-regularized quantum theory defined by the partition function $\mathcal{Z}[\tilde{\psi}]$ only diverge at phase transitions, and the limit of vanishing lattice spacing can only cancel their divergences at a fixed point along a phase transition. For instance, although the correlation length $\tilde{\xi}$ diverges all along a second order phase transition, one only obtains a finite mass $m$ \emph{via} the relation \eqref{correlationlength} at a fixed point along this phase transition. On the basis of this widely relevant example, one generally expects the presence of a fixed point to be contingent on the presence of a second order phase transition. The presence of a second order phase transition does not, however, entail the presence of a fixed point. Determining the phase structure of the partition function $\mathcal{Z}[\tilde{\psi}]$, which hopefully includes a second order transition, thus constitutes a primary goal. Depending on this phase structure, one might find multiple scaling regimes, each presumably controlled by a distinct fixed point requiring its own finite size scaling \emph{Ansatz}. 

While the finite size scaling \emph{Ansatz} \eqref{FSSAnsatz} is formulated in the limit of vanishing lattice spacing, %only holds strictly at the fixed point, 
one nevertheless expects this scaling to hold approximately for small but finite lattice spacing. % in the vicinity of the fixed point. %; that is, corrections to the scaling should be subdominant. 
Indeed, one only ever employs a finite size scaling \emph{Ansatz} at finite lattice spacing owing to the limitations imposed by numerical methods: phase transitions---and therefore fixed points---occur only in the thermodynamic limit, and, moreover, the fixed point occurs only for the relevant bare couplings precisely tuned to criticality. %which one moreover never reaches in a numerical setting, %one in fact makes use of the finite size scaling \emph{Ansatz} only in the approach to a fixed point. 
%Indeed, the utility of a finite size scaling \emph{Ansatz} is in the approach to a fixed point. 
%The necessity of running numerical simulations complicates this goal because technically phase transitions occur only in the thermodynamic limit. 
One attempts to infer the presence of phase transitions by extrapolating one's results for finite lattice extent on the basis of finite size scaling analyses of several different lattice extents. 

%For the lattice spacing to vanish while physical quantities remain finite, the discrete observable $\mathcal{O}$ entering the finite size scaling \emph{Ansatz} \eqref{FSSAnsatz} must diverge. %(If the finite size scaling \emph{FSSAnsatz} also involves taking the limit of infinite lattice extent $\tilde{L}$, then For the lattice extent to diverge while physical quantities remain finite, . . .) 
%For instance, for a scalar field $\tilde{\Phi}$ of mass $\tilde{m}$, the correlation length $\tilde{\xi}$ must diverge as the lattice spacing vanishes to yield a finite mass $m$. The correlation length $\tilde{\xi}$ of the field $\tilde{\Phi}$ only diverges at a second order phase transition of the partition function $\mathcal{Z}[\tilde{\psi}]$. (EXPLAIN further, criticality) %describing the physics of the scalar field $\tilde{\Phi}$. 
%On the basis of this widely relevant example, one generally expects the presence of a fixed point to be contingent on the presence of a second order phase transition.
%illustrates a general property of statistical mechanical models defined by partition functions: discrete quantities only diverge at phase transitions. 
%Typically, the presence of a fixed point is thus contingent on the presence of a second order phase transition 

In addition to prescribing the scaling of discrete observables towards their continuous counterparts, a finite size scaling \emph{Ansatz} finds practical utility in replacing the lattice spacing, over which one has little control, with the discrete observable $\mathcal{O}$, over which one has some control, as the quantity parametrizing the approach to the continuum description. For a constant value of the physical observable $\mathscr{O}$, the finite size scaling \emph{Ansatz} \eqref{FSSAnsatz} dictates how a different observable $\mathcal{O}'$ scales towards its continuous analogue $\mathscr{O}'$:
\begin{equation}
\mathscr{O}'=\lim_{a\rightarrow0}\mathcal{O}'a^{q'}=\lim_{\mathcal{O}\rightarrow\infty}\mathcal{O}'\left(\frac{\mathscr{O}}{\mathcal{O}}\right)^{q'/q}.
\end{equation}
%One replaces the limit of vanishing lattice spacing with the more readily controllable limit of diverging observable $\mathcal{O}$. 
% one approaches the continuum limit by letting the lattice spacing $a$ decrease to zero and the dimensionless lattice extent $\tilde{L}$ increase without bound while their product $L=\tilde{L}a$ remains finite. 
Since the physical observable $\mathscr{O}$ entering the finite size scaling \emph{Ansatz} \eqref{FSSAnsatz} is dimensionful, holding its value constant in the limit of vanishing lattice spacing $a$ or diverging discrete observable $\mathcal{O}$ requires a standard unit of measure with respect to which one measures its value.
How does one select a finite size scaling \emph{Ansatz} without already knowing how discrete observables scale into their continuous counterparts? As with the selection of a model, the selection of a finite size scaling \emph{Ansatz} essentially boils down to an exercise of guess and check in which one judges a choice of finite size scaling \emph{Ansatz} on the basis of the outcomes of statistical analyses of numerical measurements interpreted with respect to this choice. 

%Of course, the model that ultimately employs will also be influenced by which physical observables one chooses (or rather knows how) to measure. 

%One must answer the question of many physical observables are necessary to determine a trajectory. This question is intimately connected to the theory that one chooses to use to analyze data as one will typically need more observables to discern more couplings. %In this context what determines the scale $l$. The lattice spacing essentially serves this purpose. Although a priori there is no value associated to the lattice spacing, it is just an arbitrary constant in the formulation of the theory, one can consider changes in the lattice spacing. One can eventually bring the lattice spacing into contact with physical scales by making contact with known phenomena. Since the lattice in this case is still a fixed background, one is licensed in assuming that a coarse graining step increases the lattice spacing. %This will no longer be the case, however, in the context of lattice quantum gravity. Indeed, to a certain extent, this is not the case in lattice quantum field theory either, except that results bear out this assumption as being valid. 

%If I were proposing a coarse graining scheme, then the trajectories would be defined by successive steps in the coarse graining process. Even in this case, though, these trajectories would only be well defined if I ensured that my coarse graining scheme were apt. 

Equipped with a model and a finite size scaling \emph{Ansatz}, one now begins a renormalization group analysis. %Given only ensembles of configurations of the field $\tilde{\Phi}$ generated by Markov chain Monte Carlo methods, how does one proceed? 
The model provides for the delineation of renormalization group trajectories within its associated truncation $\mathfrak{t}$ of the space $\mathfrak{T}$. %corresponding to one's model. %Quantities that constitute physical observables are defined with respect to one's selected model, and 
As I observed in subsection \ref{conceptualRG}, renormalization group transformations do not change the physics, in this case, of the field $\Phi$. Accordingly, one may define renormalization group trajectories as those curves within the truncation $\mathfrak{t}$ along which this physics remains constant. %is unchanged under the action of successive renormalization group transformations. 
Since the values of the model's physical observables encode the physics of the field $\Phi$, these values---at least those to which one has access after each renormalization group transformation---should remain unchanged. To ensure that renormalization group transformations act properly, one monitors the values of physical observables under successive renormalization group transformations, checking that these values are indeed invariant. This criterion fulfills the role played by the mode decomposition of subsubsection \ref{QFTCNDST}. %To distinguish renormalization group trajectories from one another, one must consider a sufficient number of physical observables, again related to the chosen truncation. 

A physical observable of the selected model is either dimensionless or dimensionful. As I remarked in subsubsection \ref{QFTCNDST}, one requires a standard unit of measure in terms of which to determine the value of a dimensionful physical observable. In the absence of such a standard, one cannot ascribe meaning to the value of a dimensionful physical observable, let alone the constancy of this value under renormalization group transformations. %In the latter case one thus requires a standard unit of measure in which to gauge the value of the physical observable. 
In principle, and, indeed, in practice, a standard unit of measure is itself based on a dimensionful physical observable of the model---a particular physical observable that one employs in the capacity of a standard owing to its qualities as such. One might attempt to circumvent the establishment of a standard unit of measure by considering only dimensionless physical observables, including those formed by appropriate ratios of dimensionful physical observables. After all the value of a dimensionful physical observable in units of a standard of measure is a dimensionless number. If one requires, for instance, a notion of a succession of length scales, for which application of the renormalization group certainly calls, then one cannot circumvent the establishment of a standard unit of measure by only considering dimensionless physical observables. Suppose, for instance, that one measures the ratios $\ell_{1}/\ell_{2}$ and $\ell_{2}/\ell_{3}$ of three length scales $\ell_{1}$, $\ell_{2}$, and $\ell_{3}$ finding that $\ell_{1}/\ell_{2}\leq\ell_{2}/\ell_{3}$. From this information alone one cannot order the three scales $\ell_{1}$, $\ell_{2}$, and $\ell_{3}$; one requires a common standard unit of length in terms of which to compare these three scales. 

As I commented in subsection \ref{conceptualRG}, one is nevertheless warranted in supposing a standard unit of measure defined independently of one's model in certain restricted settings. For instance, I assumed the presence of a standard unit of measure defined independently of the dynamics of the field $\Phi$ in subsubsection \ref{QFTCNDST}. I justified this assumption as an extension of the fixed background structure provided by the spacetime manifold $\mathscr{M}$ with metric tensor $\mathbf{g}$. The status of a standard unit of measure is essentially the same in the current setting: the spacetime $\mathcal{M}$ provides the lattice-regularized quantum theory with the necessary fixed background structure, and the corresponding spacetime provides the continuum description with the necessary fixed background structure. There is only one slight subtlety. As I emphasized above, the value of the lattice spacing is not established in terms of a known standard unit of length. The lattice spacing can nevertheless serve as a standard unit of length in this context because its value is unaffected by the dynamics of the field $\tilde{\Phi}$. Moreover, under a renormalization group transformation, described just below, one knows how the lattice spacing changes independently of how the field $\tilde{\Phi}$ changes, so one can continue to use the lattice spacing as a standard unit of length. If the continuum description happens to coincide with a quantum theory describing actual phenomena, then one can use the correspondence established by the finite size scaling \emph{Ansatz} to infer the value of the lattice spacing in terms of a known standard unit of length. %One can eventually attempt to associate the lattice spacing with a physically defined length scale by making contact between the theory $\mathscr{T}$ and observed phenomena. 

%The presence of the fixed spacetime $\mathcal{M}$, however, serves as a stand-in for such a standard in the following sense. An implicit part of the model like that specified by the action above is the fixed spacetime $\mathscr{M}$ which one uses to define length scales. By assuming the existence of this fixed background structure, one also licenses oneself to assume the existence of an independently defined standard unit of measure. 

Next one requires a means to implement a renormalization group transformation.  %One requires a method to implement renormalization group transformations. One still aims to integrate out modes of the field $\tilde{\Phi}$. 
Typically, one employs a coarse graining scheme. Such a scheme consists of two operations implemented in concert on a configuration of the field $\tilde{\Phi}$ from an ensemble generated by Markov chain Monte Carlo simulations. One operation coarsens the lattice regularization while the other operation coarsens the degrees of freedom of the field $\tilde{\Phi}$. Many coarse graining schemes are based on a blocking procedure. Its first operation consists of dividing the spacetime $\mathcal{M}$ into blocks, effective lattice units each formed by locally merging together a small fixed number of basic lattice units according to a prescribed rule. The second operation consists of assigning an effective value of the field $\tilde{\Phi}$ to each block computed from a prescribed function of the values of the field $\tilde{\Phi}$ at each basic lattice unit. 
The blocking procedure is intended to sum out degrees of freedom on scales smaller than that of the block. Since the spacetime $\mathcal{M}$ is fixed, and thus independent of the dynamics of the field $\tilde{\Phi}$, one is licensed in assuming that each edge of the lattice possesses a well-defined length so that blocked lattice edges do indeed correspond to larger lengths. % merged lattice units do correspond to a larger lattice spacing. %Since one cannot simply integrate out modes as in the continuum case, 
After implementing this pair of operations, one rescales all dimensionful quantities by the ratio of the block scale to the basic lattice unit scale. This rescaling ensures that the coarsened spacetime has the same physical scale as the original spacetime.
%one continues to consider the system as being of the same physical size. 
One may not employ just any coarse graining scheme because not all coarse graining schemes correspond to renormalization group transformations. One must ensure that one's coarse graining scheme satisfies the essential feature of a renormalization group transformation, that of preserving the physics of the field $\tilde{\Phi}$. Such a coarse graining scheme is deemed apt. Within the setting of numerical analysis, one allows for this preservation to be only approximate, that is, sufficiently accurate for one's purposes. % does not deviate from one trajectory to another. %It becomes crucial that one identify physical quantities to hold fixed along the trajectory for the constancy of the physical quantities will serve to define the trajectory. 
%Only coarse graining procedures that maintain (usually approximately) the constancy of physical observables are appropriate.  

Now one extracts the renormalization group flows of the couplings $c_{j}$. Starting from an ensemble of configurations of the field $\tilde{\Phi}$, one iterates the renormalization group transformation on each configuration. After each transformation one determines the renormalized values of the couplings $c_{j}$ by measuring the appropriate discrete observables and employing the finite size scaling \emph{Ansatz} to connect their values to the appropriate physical observables. % applying the finite size scaling $\emph{Ansatz}$. 
One thus traces the renormalization group flows as the ultraviolet scale $\tilde{\ell}_{\mathrm{UV}}$ is incrementally increased, thereby inferring the Lagrangian $\mathscr{L}_{\mathscr{T}}^{(\mathrm{E})}[\Phi]$ at each step. Although one cannot associate a physical length scale to the successive intervals of scales being probed, one determines all of the structure induced on the space $\mathfrak{T}$ within the truncation $\mathfrak{t}$ by the renormalization group transformation. %the form of the renormalization group flows and all of the physical information that comes with that. 
From each ensemble of configurations of the field $\tilde{\Phi}$, both the original and the coarse grained, one thus determines a particular part of the continuum description given by a theory $\mathscr{T}$ effective on some interval of scales. The ensembles of configurations of the field $\tilde{\Phi}$ obtained one from another by the coarse graining operation give rise to a renormalization group trajectory of theories $\mathscr{T}^{(l)}$. If the bare couplings $\tilde{c}_{j}$ of the Lagrangian $\mathcal{L}_{\mathrm{cl}}^{(\mathrm{E})}[\tilde{\Phi}]$ are appropriately tuned to criticality, then this renormalization group trajectory may contain a fixed point. %If any continuum description of the lattice-regularized quantum theory exists, then one obtains access the this continuum on some interval of scales by analyzing one particular ensemble. %(Make clearer that each ensemble is for different values of the bare couplings, so different values of the bare couplings let one access different regimes of the continuum description)

Since one only introduced the lattice regularization to define nonperturbatively the quantum theory of the field $\Phi$, one is particularly interested in searching for fixed points of the renormalization group transformation at which one can remove the lattice spacing---the ultraviolet regulator---or the lattice extent---the infrared regulator. % regulating scales at a fixed point while physical quantities remain finite.
%within the quantum theory defined by the partition function $\mathcal{Z}[\tilde{\psi}]$. %To approach a fixed point along a particular renormalization group trajectory, 
Two limitations of numerical analysis complicate this search. First, as I previously commented, one cannot attain the limit of vanishing lattice spacing or divergent lattice extent. One thus attempts to infer the presence of a fixed points through the behavior of renormalization group trajectories that (presumably) pass close to this fixed point. %one always works at finite lattice spacing $a$ and finite lattice extent $\tilde{L}$. 
Second, one can only implement renormalization group transformations as coarse graining operations, which limits one to constructing renormalization group trajectories % In the numerical analysis of ensembles of configurations of the field $\tilde{\Phi}$, the coarse graining operation is the only renormalization group transformation that one can implement, so one can only study the renormalization group trajectories that it generates, namely renormalization group trajectories 
for which the ultraviolet scale $\tilde{\ell}_{\mathrm{UV}}$ is incrementally increased. How does one determine the presence of an ultraviolet or an infrared fixed point from this restricted information? 

To search for an ultraviolet fixed point, % is conceptually straightforward. Once one has constructed these renormalization group trajectories, 
one simply follows the constructed renormalization group trajectories in the direction opposite to that induced by the coarse graining operation, in the direction of decreasing ultraviolet scale $\tilde{\ell}_{\mathrm{UV}}$. % them backwards in the direction of decreasing $\tilde{\ell}_{\mathrm{UV}}$. 
If any of these renormalization group trajectories approaches a second order phase transition when so followed, then there might exist an ultraviolet fixed point along this transition. In particular, for critical---specifically tuned---values of some (or possibly all) of the bare couplings $\tilde{c}_{j}$ of the Lagrangian $\mathcal{L}_{\mathrm{cl}}^{(\mathrm{E})}[\tilde{\Phi}]$, % along a second order phase transition.
one should find that the lattice spacing necessarily approaches zero as one follows this renormalization group trajectory towards the second order phase transition. The condition of constant physics, which delineated the renormalization group trajectory in the first place, entails the vanishing of the lattice spacing.
% to those at the second order phase transition and tune the relevant couplings to criticality. 
%To determine the presence of a fixed point, one studies the properties of renormalization group flows in the vicinity of any second order phase transitions.  If there is a renormalization group trajectory with the following property, then one has identified an ultraviolet fixed point. The coarse graining operation induces a direction on the renormalization group trajectory that it generates. One wants to follow this renormalization group trajectory in the opposite direction towards smaller values of $\tilde{\ell}_{\mathrm{UV}}$. 
%To determine the nature of a fixed point, one studies the properties of renormalization group flows in its vicinity.  
%In particular, if one finds that the lattice spacing decreases towards zero as one moves along a renormalization group trajectory towards a fixed point, then that fixed point is an ultraviolet fixed point. 

To search for an infrared fixed point, one requires a more involved procedure. Although iterating the coarse graining operation yields the physics that is most relevant on large scales bounded by the scale $\tilde{\ell}_{\mathrm{IR}}$, the coarse graining operation, with its subsequent rescaling of all dimensionful quantities, preserves the lattice extent $L$. %To find an infrared fixed point, one requires a slightly different analysis because the rescaling following the coarse graining operation ensures that the scale $\ell_{\mathrm{IR}}$ remains fixed. 
%Subsequent iterations of the coarse graining operation do yield the physics that is most relevant on large scales up to the scale $\ell_{\mathrm{IR}}$. 
To probe renormalization group trajectories along which the scale $\tilde{\ell}_{\mathrm{IR}}$ increases without bound, one must consider lattice-regularized spacetimes of successively larger extent $L$. If any renormalization group trajectories so constructed approach a second order phase transition, then there might exist an infrared fixed point along this transition. Again, for critical values of the bare couplings $\tilde{c}_{j}$ of the Lagrangian $\mathcal{L}_{\mathrm{cl}}^{(\mathrm{E})}[\tilde{\Phi}]$, one should find that the lattice extent necessarily approaches infinity as one follows this renormalization group trajectory towards the second order phase transition. The condition of constant physics, which delineated the renormlization group trajectory in the first place, again entails the divergence of the lattice extent. %If one finds that the lattice extent $\tilde{L}$ increases towards infinity as one moves along a renormalization group trajectory towards a fixed point, then that fixed point is an infrared fixed point. %In this sense if a fixed point exists in the ultraviolet, then one can formally remove the regulator $\mathfrak{l}_{\mathrm{UV}}$, in this case the lattice spacing $a$, and, if a fixed point exists in the infrared, then one can formally remove the regulator $\mathfrak{l}_{\mathrm{IR}}$, in this case the lattice extent $\tilde{L}$. If either or both of these fixed points is absent, then one may nevertheless be able to obtain a continuum description effective over a nontrivial interval of scales, even including nonperturbative physics. %In this case one's model of the continuum theory would already contain the contributions from counterterms. Since one infers the values of physical observables from numerical measurements, one necessarily obtains finite renormalized values. 

%Approaching the continuum limit typically requires that one tune the couplings $c_{j}$ of the Lagrangian $\mathcal{L}_{\mathrm{cl}}[\tilde{\Phi}]$ to their values corresponding to a second order phase transition of the statistical mechanical model defined by the partition function $\mathcal{Z}[\tilde{\psi}]$. 

%When one starts to analyze such numerical data, %for instance, an ensemble of field configurations, 
%one knows neither how to take the continuum limit nor what the continuum limit is (assuming that it exists). %In analyzing numerical measurements, 
%If one does find a second order phase transition, then one must first make assumptions about how to take the continuum limit and what the continuum limit is, revising these assumptions in light of the outcomes of statistical analyses of numerical data. 

%This is the procedure that one follows in an attempt to derive a continuum description from a lattice-regularized quantum theory.

\subsubsection{Quantum field theory of continuous dynamical spacetime}\label{QG}

\paragraph{Definition}

Consider next the quantum theory of a symmetric spacetime metric tensor $\mathbf{g}$. % of the propagation of the dynamical field $\Phi$ on a spacetime $\mathcal{M}$ with dynamical metric tensor $\mathbf{g}$ governed by the quantum effective action
%\begin{equation}
%S_{\mathscr{T}}[\Phi,\mathbf{g}]=\int_{\mathcal{M}}\mathrm{d}^{4}x\sqrt{-g}\mathcal{L}[\Phi,\mathbf{g}].
%\end{equation}
As in subsubsection \ref{QFTCNDST}, I aim to determine the Lagrangian $\mathscr{L}_{\mathscr{T}}[\mathbf{g}]$ that provides an effective description of the dynamics of the metric tensor $\mathbf{g}$ within the interval of scales $(\ell_{\mathrm{UV}},\ell_{\mathrm{IR}})$. The quantum effective action $S_{\mathscr{T}}[\mathbf{g}]$---the spacetime integral of the Lagrangian $\mathscr{L}_{\mathscr{T}}[\mathbf{g}]$---generically assumes the form 
\begin{equation}\label{metriceffectiveactionderexp}
S_{\mathscr{T}}[\mathbf{g}]=\int_{\mathscr{M}}\mathrm{d}^{d+1}x\sqrt{-g}\left[c_{0}+c_{1}R+c_{2}R^{2}+c_{3}R_{ab}R^{ab}+c_{4}R_{abcd}R^{abcd}+c_{5}R^{3}+\cdots\right],
\end{equation}
an expansion in derivatives of the metric tensor $\mathbf{g}$. The quantities $1$, $R$, $R^{2}$, $R_{ab}R^{ab}$, $R_{abcd}R^{abcd}$, $R^{3}$, $\ldots$---all of the scalar functionals that one can construct from the metric tensor, its Riemann tensor, and covariant derivatives of its Riemann tensor---constitute a complete basis for this expansion.\footnote{If one considers a model that requires more data than that contained within a symmetric metric tensor to specify the geometry of spacetime, then one must generalize appropriately the basis of scalar functionals.} %(comment on what spacetime/metric are used in the effective action---the expectation value in the quantum state defined by the path integral)
Arriving at the Lagrangian $\mathscr{L}_{\mathscr{T}}[\mathbf{g}]$ is, however, neither conceptually nor technically straightforward, primarily owing to the challenge of defining length scales---let alone a particular interval of length scales---within the quantum theory of the metric tensor $\mathbf{g}$. Although any particular metric tensor $\mathbf{g}$ defines length scales, the quantum theory inextricably involves superpositions of metric tensors $\mathbf{g}$. I now attempt nevertheless to determine the Lagrangian $\mathscr{L}_{\mathscr{T}}[\mathbf{g}]$. 

Given a classical theory $\mathscr{T}_{\mathrm{cl}}$ of a metric tensor $\mathbf{g}$ specified by the Lagrangian $\mathscr{L}_{\mathrm{cl}}[\mathbf{g}]$, one formally computes a transition amplitude $\mathscr{A}[\gamma]$ in the quantum theory as the path integral
\begin{equation}\label{QFT3state}
\mathscr{A}[\gamma]=\int_{\mathbf{g}|_{\partial\mathscr{M}}=\gamma}\mathrm{d}\mu(\mathbf{g})\,e^{iS_{\mathrm{cl}}[\mathbf{g}]/\hbar}.
\end{equation}
The metric tensor $\gamma$ induced on the boundary $\partial\mathscr{M}$ of the spacetime manifold $\mathscr{M}$ by its metric tensor $\mathbf{g}$ specifies the transition amplitude $\mathscr{A}[\gamma]$. The expression \eqref{QFT3state} is an instruction to integrate over all metric tensors $\mathbf{g}$ satisfying the boundary condition $\mathbf{g}|_{\partial\mathscr{M}}=\gamma$, weighting each by the product of the measure $\mathrm{d}\mu(\mathbf{g})$ and the exponential $e^{iS_{\mathrm{cl}}[\mathbf{g}]/\hbar}$. There is still some freedom in the definition of the measure $\mathrm{d}\mu(\mathbf{g})$. One usually still designs the measure $\mathrm{d}\mu(\mathbf{g})$ to respect any invariances of the theory $\mathscr{T}_{\mathrm{cl}}$ so that one does not introduce additional degrees of freedom and to eliminate any redundancies of description of the theory $\mathscr{T}_{\mathrm{cl}}$ %---diffeomorphisms in this case if one has coordinatized the spacetime manifold $\mathscr{M}$---
so that one integrates only over physically distinct metric tensors $\mathbf{g}$.%How is one to make sense of scales for this path integral?

One expects the path integral in equation \eqref{QFT3state} to again be formally divergent, so one must introduce a regularization to render it well-defined. In subsubsection \ref{QFTCNDST} I introduced an ultraviolet regulator $\mathfrak{l}_{\mathrm{UV}}$ and an infrared regulator $\mathfrak{l}_{\mathrm{IR}}$ for this purpose. The scale $\mathfrak{l}_{\mathrm{UV}}$ served as a regulator by providing a lower bound on the length scale of field excitations, and the scale $\mathfrak{l}_{\mathrm{IR}}$ served as a regulator by providing an upper bound on the length scale of field excitations. I identified the scale of field excitations with respect to a fixed spacetime manifold with definite metric tensor, a background structure in the formulation of the quantum theory. The quantum theory of a metric tensor $\mathbf{g}$ is not, however, defined in the presence of such a background structure, so one cannot identify the scale of field excitations in the same manner.\footnote{Within the functional renormalization group approach to the construction of a quantum theory of gravity, one performs a mode decomposition of the metric tensor $\mathbf{g}$ using the background field method. One introduces an arbitrary but fixed metric tensor $\mathbf{g}_{\mathrm{ref}}$ as a background structure with respect to which the mode decomposition of the metric tensor $\mathbf{g}$ is defined. After integrating out modes of the metric tensor $\mathbf{g}$ in the path integration, the metric tensor $\mathbf{g}_{\mathrm{ref}}$ is dynamically adjusted so that all results are covariant with respect to the sum of the metric tensors. One argues that this procedure provides a physically meaningful prescription for identifying scales \cite{MN&MR,MR}.}

%One might nevertheless hope to define the Lagrangian $\mathscr{L}_{\mathscr{T}}[\mathbf{g}]$ in analogy to the procedure of subsubsection \ref{QFTCNDST}. 
Since one cannot perform a mode decomposition of each metric tensor $\mathbf{g}$, one requires a different approach to regularization. %Since a mode decomposition of each metric tensor $\mathbf{g}$ is problematic, 
Mimicking the prescription of equation \eqref{QFT1pathintegral'}, one might attempt to regularize 
%define the Lagrangian $\mathscr{L}_{\mathscr{T}}[\mathbf{g}]$ . To regularize 
the path integration in equation \eqref{QFT3state} by introducing the integral kernel $\Omega_{(\mathfrak{l}_{\mathrm{UV}},\mathfrak{l}_{\mathrm{IR}})}(\mathbf{g})$. I wish to consider this approach to illustrate the profound difficulties that one must overcome. %As in equation \eqref{QFT1regularizedmeasure} 
Suppose that one constructs a candidate integral kernel $\Omega_{(\mathfrak{l}_{\mathrm{UV}},\mathfrak{l}_{\mathrm{IR}})}(\mathbf{g})$. One intends the integral kernel $\Omega_{(\mathfrak{l}_{\mathrm{UV}},\mathfrak{l}_{\mathrm{IR}})}(\mathbf{g})$ to give rise to a regularized measure
\begin{equation}\label{QFT3regularizedmeasure}
\mathrm{d}\mu_{(\mathfrak{l}_{\mathrm{UV}},\mathfrak{l}_{\mathrm{IR}})}(\mathbf{g})=\mathrm{d}\mu(\mathbf{g})\,\Omega_{(\mathfrak{l}_{\mathrm{UV}},\mathfrak{l}_{\mathrm{IR}})}(\mathbf{g}),
\end{equation}
which assigns zero weight to excitations of each metric tensor $\mathbf{g}$ on scales $\ell\in(0,\mathfrak{l}_{\mathrm{UV}})\cup(\mathfrak{l}_{\mathrm{IR}},\infty)$. How does one determine whether or not the candidate integral kernel $\Omega_{(\mathfrak{l}_{\mathrm{UV}},\mathfrak{l}_{\mathrm{IR}})}(\mathbf{g})$ functions as intended?

First of all, the integral kernel $\Omega_{(\mathfrak{l}_{\mathrm{UV}},\mathfrak{l}_{\mathrm{IR}})}(\mathbf{g})$ must give rise to a regularized transition amplitude
\begin{equation}\label{QFT3statereg}
\bar{\mathscr{A}}[\gamma]=\int_{\mathbf{g}|_{\partial\mathscr{M}}=\gamma}\mathrm{d}\mu_{(\mathfrak{l}_{\mathrm{UV}},\mathfrak{l}_{\mathrm{IR}})}(\mathbf{g})\,e^{iS_{\mathrm{cl}}[\mathbf{g}]/\hbar}
\end{equation}
%, computed as in equation \eqref{QFT3state} but for the measure \eqref{QFT3regularizedmeasure}, 
for which the expectation values of all physical observables %computed in the quantum state \eqref{QFT3statereg} 
are finite. As in subsubsection \ref{QFTCNDST}, if the measure $\mathrm{d}\mu_{(\mathfrak{l}_{\mathrm{UV}},\mathfrak{l}_{\mathrm{IR}})}(\mathbf{g})$ respects the invariances of the theory $\mathscr{T}_{\mathrm{cl}}$, then in principle any functional $\mathscr{O}[\mathbf{g}]$ of the metric tensor $\mathbf{g}$ that also respects these invariances counts as a physical observable, and, if the measure $\mathrm{d}\mu_{(\mathfrak{l}_{\mathrm{UV}},\mathfrak{l}_{\mathrm{IR}})}(\mathbf{g})$ does not respect the invariances of the theory $\mathscr{T}_{\mathrm{cl}}$, owing either to the unregularized measure $\mathrm{d}\mu(\mathbf{g})$ or the integral kernel $\Omega_{(\mathfrak{l}_{\mathrm{UV}},\mathfrak{l}_{\mathrm{IR}})}(\mathbf{g})$, then in principle any functional $\mathscr{O}[\mathbf{g}]$ of the metric tensor $\mathbf{g}$ that respects the remaining subset of these invariances counts as a physical observable. One generally expects the expectation value of a physical observable in the quantum state \eqref{QFT3statereg} to depend on the regularizing scales $\mathfrak{l}_{\mathrm{UV}}$ and $\mathfrak{l}_{\mathrm{IR}}$ that enter into the definition of the integral kernel $\Omega_{(\mathfrak{l}_{\mathrm{UV}},\mathfrak{l}_{\mathrm{IR}})}(\mathbf{g})$.

Furthermore, the interval of scales dynamically explored by the superposition of metric tensors $\mathbf{g}$ within the transition amplitude \eqref{QFT3statereg} must be that delimited by the regularizing scales $\mathfrak{l}_{\mathrm{UV}}$ and $\mathfrak{l}_{\mathrm{IR}}$. How does one ascertain the interval of scales dynamically so explored? There is only one recourse: one must compute the expectation values of sufficiently many physical observables to determine the interval of scales represented within these expectation values. %to ascertain the interval of scales dynamically explored by the superposition of metric tensors $\mathbf{g}$ within the regularized transition amplitude $\bar{\mathscr{A}}[\gamma]$. % How can one ascertain the interval of scales dynamically explored by the superposition of metric tensors $\mathbf{g}$ within a transition amplitude $\mathscr{A}[\gamma]$? There is only one recourse: one must compute the expectation values of sufficiently many physical observables and determine the range of scales represented within these expectation values. For the transition amplitude $\mathscr{A}[\gamma]$ computed with the regularized measure \eqref{QFT3regularizedmeasure}, % regularized by the integral kernel $\Omega_{(\mathfrak{l}_{\mathrm{UV}},\mathfrak{l}_{\mathrm{IR}})}(\mathbf{g})$, 
This amounts to checking the relevance of the length scales $\mathfrak{l}_{\mathrm{UV}}$ and $\mathfrak{l}_{\mathrm{IR}}$, introduced within the integral kernel $\Omega_{(\mathfrak{l}_{\mathrm{UV}},\mathfrak{l}_{\mathrm{IR}})}(\mathbf{g})$, 
for regularizing the path integration in equation \eqref{QFT3state}. In subsubsection \ref{QFTCNDST} the mode decomposition of the field $\Phi$ made clear the relevance of the length scales $\mathfrak{l}_{\mathrm{UV}}$ and $\mathfrak{l}_{\mathrm{IR}}$ for regularizing the path integration in equation \eqref{QFT1state}. %was clear owing to the mode decomposition because these length scales and those emerging from the dynamics of the field $\Phi$ were measured with respect to a fixed background structure. %One must now ascertain the physical relevance of the length scales $\mathfrak{l}_{\mathrm{UV}}$ and $\mathfrak{l}_{\mathrm{IR}}$. 

%There is yet another issue that one must address: 
These checks of the functioning of the integral kernel $\Omega_{(\mathfrak{l}_{\mathrm{UV}},\mathfrak{l}_{\mathrm{IR}})}(\mathbf{g})$ require the specification of a standard unit of measure in terms of which the scales represented within expectation values of physical observables and the expectation values of dimensionful physical observables are determined. Within the data specified by the metric tensor $\gamma$ induced on the boundary $\partial\mathscr{M}$, one may include information about a candidate standard unit of measure. %If one does specify boundary data, then one assumes the presence of a classical regime. If one does specify boundary data---for instance, the fixed metric tensors $\gamma_{i}$ and $\gamma_{f}$ on initial and final spacelike hypersurfaces---then one can import classically defined standards of reference defined by these boundary data. 
This boundary data constitutes classically specifiable information. The inclusion of appropriate boundary terms in the action $S_{\mathrm{cl}}[\mathbf{g}]$, encoding the boundary data, serves to establish the dynamical relationship between the scales defined by the boundary data and the scales emerging from the transition amplitude. This relationship endows the boundary data with physical meaning, and one must ascertain whether or not this relationship ascribes the desired physical meaning to a candidate standard unit of measure. One may, however, choose to work in a cosmological setting in which case one does not specify any boundary data. %\footnote{One may well be of the opinion that one the first case makes sense. This is indeed the setting for all unambiguous experimental tests of quantum theory. Nevertheless, the universe is by definition a closed system. I attempt to make sense of both situations in the following.} 
%If one does not specify boundary data---for instance, when working in a cosmological setting---then 
There are then no fixed reference scales, only scales emerging dynamically from the transition amplitude \eqref{QFT3state}. %The only scales present are those emerging dynamically from the transition amplitude. 
In this situation one must employ as a standard unit of measure one of these dynamically emergent scales. Isolating a dynamically emergent scale having the characteristics that one usually demands of a standard unit of measure %that  such a scale that behaves as one would expect a standard unit of measure to behave 
might prove a quite nontrivial problem. 

%One attempts to define such a chosen reference in accord with the criteria that metrologists use to define such standards as the meter, the second, and the kilogram.

The integral kernel $\Omega_{(\mathfrak{l}_{\mathrm{UV}},\mathfrak{l}_{\mathrm{IR}})}(\Phi)$ that I introduced in subsubsection \ref{QFTCNDST} automatically preserved the physics of the field $\Phi$ on scales $\ell\in(\mathfrak{l}_{\mathrm{UV}},\mathfrak{l}_{\mathrm{IR}})$. Its definition \emph{via} the mode decomposition of the field $\Phi$ ensured this property of the integral kernel $\Omega_{(\mathfrak{l}_{\mathrm{UV}},\mathfrak{l}_{\mathrm{IR}})}(\Phi)$. Supposing that the integral kernel $\Omega_{(\mathfrak{l}_{\mathrm{UV}},\mathfrak{l}_{\mathrm{IR}})}(\mathbf{g})$ passes the checks just discussed, how can one ascertain whether or not the integral kernel $\Omega_{(\mathfrak{l}_{\mathrm{UV}},\mathfrak{l}_{\mathrm{IR}})}(\mathbf{g})$ preserves the physics of the metric tensor $\mathbf{g}$ on scales $\ell\in(\mathfrak{l}_{\mathrm{UV}},\mathfrak{l}_{\mathrm{IR}})$? There is apparently no recourse: one cannot perform a mode decomposition of the metric tensor $\mathbf{g}$, and one cannot compute expectation values of physical observables in the absence of the regularization provided by the integral kernel $\Omega_{(\mathfrak{l}_{\mathrm{UV}},\mathfrak{l}_{\mathrm{IR}})}(\mathbf{g})$. %, and the expectation values of physical observables computed in the presence of the regularization depend on the regularizing scales $\mathfrak{l}_{\mathrm{UV}}$ and $\mathfrak{l}_{\mathrm{IR}}$. 
Accordingly, one must accept the regularization provided by the integral kernel  $\Omega_{(\mathfrak{l}_{\mathrm{UV}},\mathfrak{l}_{\mathrm{IR}})}(\mathbf{g})$ as part of the definition of the quantum theory of the metric tensor $\mathbf{g}$. Only in retrospect can one ultimately judge the choice of integral kernel  $\Omega_{(\mathfrak{l}_{\mathrm{UV}},\mathfrak{l}_{\mathrm{IR}})}(\mathbf{g})$.

Assuming success in constructing the integral kernel $\Omega_{(\mathfrak{l}_{\mathrm{UV}},\mathfrak{l}_{\mathrm{IR}})}(\mathbf{g})$, I am now prepared to define the regularized Lagrangian $\bar{\mathscr{L}}_{\bar{\mathscr{T}}}[\mathbf{g}]$ effective on the interval of scales $(\ell_{\mathrm{UV}},\ell_{\mathrm{IR}})$. The regularized quantum effective action $\bar{S}_{\bar{\mathscr{T}}}[\mathbf{g}]$---the spacetime integral of the Lagrangian $\bar{\mathscr{L}}_{\bar{\mathscr{T}}}[\mathbf{g}]$---is defined through the relation 
%Once one has settled on an integral kernel $\Omega_{(\mathfrak{l}_{\mathrm{UV}},\mathfrak{l}_{\mathrm{IR}})}(\mathbf{g})$, one then defines the regularized Lagrangian $\bar{\mathscr{L}}_{\bar{\mathscr{T}}}[\mathbf{g}]$ effective on the interval of scales $(\ell_{\mathrm{UV}},\ell_{\mathrm{IR}})$ through the introduction of another integral kernel:
\begin{equation}\label{QFT3statereg}
e^{i\bar{S}_{\bar{\mathscr{T}}}[\mathbf{g}]/\hbar}=\int_{\mathbf{g}|_{\partial\mathscr{M}}=\gamma}\mathrm{d}\mu_{(\mathfrak{l}_{\mathrm{UV}},\mathfrak{l}_{\mathrm{IR}})}(\mathbf{g})\,\Omega_{(\ell_{\mathrm{UV}},\ell_{\mathrm{IR}})}(\mathbf{g})\,e^{iS_{\mathrm{cl}}[\mathbf{g}]/\hbar}.
\end{equation}
%Thus far, however, I have only introduced some formal notation. In subsubsection \ref{QFTCNDST} I defined the integral kernel $\Omega_{(\ell_{\mathrm{UV}},\ell_{\mathrm{IR}})}(\Phi)$ in terms of the mode decomposition of the field $\Phi$. Since this option is not available now, one must proceed differently. 
for the similarly constructed integral kernel $\Omega_{(\ell_{\mathrm{UV}},\ell_{\mathrm{IR}})}(\mathbf{g})$. %One takes the relation \eqref{QFT3statereg} essentially as a definition of the regularized quantum theory of the metric tensor $\mathbf{g}$. 
One can now ascertain whether or not the integral kernel $\Omega_{(\ell_{\mathrm{UV}},\ell_{\mathrm{IR}})}(\mathbf{g})$ preserves the (regularized) physics of the metric tensor $\mathbf{g}$ on the interval of scales $(\ell_{\mathrm{UV}},\ell_{\mathrm{IR}})$. The integral kernel $\Omega_{(\ell_{\mathrm{UV}},\ell_{\mathrm{IR}})}(\mathbf{g})$ must satisfy the analogue of condition \eqref{QFT1condition}, namely
\begin{equation}
\int_{\mathbf{g}|_{\partial\mathscr{M}}=\gamma}\mathrm{d}\mu_{(\mathfrak{l}_{\mathrm{UV}},\mathfrak{l}_{\mathrm{IR}})}(\mathbf{g})\,\Omega_{(\ell_{\mathrm{UV}},\ell_{\mathrm{IR}})}(\mathbf{g})=1,
\end{equation}
resulting from the requirement that one may compute the regularized transition amplitude \eqref{QFT3statereg} also from the action $\bar{S}_{\bar{\mathscr{T}}}[\mathbf{g}]$:
\begin{equation}\label{QFT3statereg2}
\bar{\mathscr{A}}[\gamma]=\int_{\mathbf{g}|_{\partial\mathscr{M}}=\gamma}\mathrm{d}\mu_{(\ell_{\mathrm{UV}},\ell_{\mathrm{IR}})}(\mathbf{g})\,e^{i\bar{S}_{\bar{\mathscr{T}}}[\mathbf{g}]/\hbar}.
\end{equation}
Moreover, the expectation value of a physical observable computed in the quantum state \eqref{QFT3statereg} must agree with the expectation value of that physical observable computed in the quantum state \eqref{QFT3statereg2}. If the integral kernel $\Omega_{(\ell_{\mathrm{UV}},\ell_{\mathrm{IR}})}(\mathbf{g})$ meets these criteria, then one can be reasonably confident in the physical relevancy of the integral kernel $\Omega_{(\mathfrak{l}_{\mathrm{UV}},\mathfrak{l}_{\mathrm{IR}})}(\mathbf{g})$ for regularizing the path integration in equation \eqref{QFT3state}.

The construction of an appropriate integral kernel $\Omega_{(\mathfrak{l}_{\mathrm{UV}},\mathfrak{l}_{\mathrm{IR}})}(\mathbf{g})$, the identification of physical observables, the computation of their expectation values, and the establishment of a standard unit of measure are all presumably quite technically challenging problems. To define nonperturbatively the quantum theory of the metric tensor $\mathbf{g}$ using path integral techniques, these are the difficulties that one must overcome. %Moreover, one must have sufficient physical insight into the physical observables to understand what scales are represented within their dynamics. 

\paragraph{Renormalization}

The Lagrangian $\bar{\mathscr{L}}_{\bar{\mathscr{T}}}[\mathbf{g}]$ still depends on the regulators $\mathfrak{l}_{\mathrm{UV}}$ and $\mathfrak{l}_{\mathrm{IR}}$. One can obtain the Lagrangian $\mathscr{L}_{\mathscr{T}}[\mathbf{g}]$ as in subsubsection \ref{QFTCNDST}: renormalize the Lagrangian $\bar{\mathscr{L}}_{\bar{\mathscr{T}}}[\mathbf{g}]$ \emph{via} the addition of counterterms. In particular, the regularized action $\bar{S}_{\bar{\mathscr{T}}}[\mathbf{g}]$ has the form of equation \eqref{metriceffectiveactionderexp} for couplings $\bar{c}_{j}$ that depend on the regulators $\mathfrak{l}_{\mathrm{UV}}$ and $\mathfrak{l}_{\mathrm{IR}}$. To match the predictions of the regularized theory $\bar{\mathscr{T}}$ to the outcomes of experimental measurements, one adds an appropriate counterterm for each term in the Lagrangian $\bar{\mathscr{L}}_{\bar{\mathscr{T}}}[\mathbf{g}]$. The couplings $\bar{c}_{j}$ of the Lagrangian $\bar{\mathscr{L}}_{\bar{\mathscr{T}}}[\mathbf{g}]$ combine with the couplings $\hat{c}_{j}$ of the counterterms to yield the renormalized couplings $c_{j}=\bar{c}_{j}+\hat{c}_{j}$ of the Lagrangian $\mathscr{L}_{\mathscr{T}}[\mathbf{g}]$.

Starting from the Lagrangian $\mathscr{L}_{\mathscr{T}}[\mathbf{g}]$, one can apply renormalization group transformations to obtain the Lagrangians $\mathscr{L}_{\mathscr{T}'}[\mathbf{g}]$ effective on other intervals of scales. Knowledge of the Lagrangians $\mathscr{L}_{\mathscr{T}}[\mathbf{g}]$ is again tantamount to knowledge of the renormalization group flows of the couplings $c_{j}$ parametrizing the space $\mathfrak{T}$ of theories $\mathscr{T}$ of the metric tensor $\mathbf{g}$. Consider in particular a renormalization group transformation that incrementally increases the ultraviolet scale $\mathfrak{l}_{\mathrm{UV}}$ and leaves fixed the infrared scale $\mathfrak{l}_{\mathrm{IR}}$. Since one has determined how to construct the integral kernels $\Omega(\mathbf{g})$, one may proceed as in subsubsection \ref{QFTCNDST}. One implements this renormalization group transformation by inserting the integral kernel $\Omega_{(\ell_{\mathrm{UV}}+\delta\ell,\ell_{\mathrm{IR}})}(\mathbf{g})$ into the path integral:
\begin{equation}
e^{iS_{\mathscr{T}'}[\mathbf{g}]/\hbar}=\int_{\mathbf{g}|_{\partial\mathscr{M}}=\gamma}\mathrm{d}\mu_{(\ell_{\mathrm{UV}},\ell_{\mathrm{IR}})}(\mathbf{g})\,\Omega_{(\ell_{\mathrm{UV}}+\delta\ell,\ell_{\mathrm{IR}})}(\mathbf{g})\,e^{iS_{\mathscr{T}}[\mathbf{g}]/\hbar}.
\end{equation}
This renormalization group transformation automatically maps the theory $\mathscr{T}$ into the theory $\mathscr{T}'$ describing the same physics: the integral kernel $\Omega_{(\ell_{\mathrm{UV}}+\delta\ell,\ell_{\mathrm{IR}})}(\mathbf{g})$ precisely selects modes within the interval of scales $(\ell_{\mathrm{UV}},\ell_{\mathrm{UV}}+\delta\ell)$ to be integrated out, leaving fixed the modes within the interval of scales $(\ell_{\mathrm{UV}}+\delta\ell,\ell_{\mathrm{IR}})$. One now extracts the renormalization group flows of the couplings $c_{j}$ by iterating the renormalization group transformation.

\subsubsection{Quantum field theory of lattice-regularized dynamical spacetime}\label{QGlat}

\paragraph{Definition}

As I commented in subsubsection \ref{QG}, one must regularize the path integration in equation \eqref{QFT3state} to render it well-defined. I argued that implementing this regularization while continuing to allow for dynamics on the continuum of scales between $\ell_{\mathrm{UV}}$ and $\ell_{\mathrm{IR}}$ is exceedingly difficult. One might thus turn to a lattice regularization. Such a regularization introduces ultraviolet and infrared regulators $\mathfrak{l}_{\mathrm{UV}}$ and $\mathfrak{l}_{\mathrm{IR}}$ simply by construction---albeit at \emph{a priori} unknown physical scales---at the expense of allowing for dynamics only on a discretuum of scales. One's task then consists in attempting to determine the continuum description of one's lattice-regularized quantum theory. %In this setting one finds that all of the techniques discussed in the subsubsections come together. 

Consider finally the quantum theory of lattice-regularized spacetime $\mathcal{M}$. The type of lattice regularization, including the assignment of a lattice spacing $a$, and the adjacencies of lattice constituents determine the metrical structure of a spacetime $\mathcal{M}$. As in subsubsection \ref{QFTLRNDST} the value of the lattice spacing is \emph{a priori} arbitrary. The lattice regularization thus itself encodes the field degrees of freedom. %Unlike in subsubsection \ref{QFTLRNDST} the lattice spacing cannot now serve as the chosen standard of length. 
How do the analyses of subsubsections \ref{QFTCNDST}, \ref{QFTLRNDST}, and \ref{QG} carry over to this setting? One can retain much of the analysis of subsubsection \ref{QFTLRNDST}, but one must now make accommodations for the difficulties encountered in subsubsection \ref{QG}. 

Starting from a classical theory $\mathscr{T}_{\mathrm{cl}}$ of the symmetric spacetime metric tensor $\mathbf{g}$ specified by a Lagrangian $\mathscr{L}_{\mathrm{cl}}[\mathbf{g}]$, one constructs a corresponding discrete Lagrangian $\mathcal{L}_{\mathrm{cl}}[\mathcal{M}]$ suited to the chosen lattice regularization. %The metrical structure of any particular spacetime $\mathcal{M}$ is determined by the connectivities of its lattice constituents and the assignment of lengths (in units of a lattice spacing) to their edges. 
As in subsubsection \ref{QFTLRNDST} the discrete classical action $\mathcal{S}_{\mathrm{cl}}[\mathcal{M}]$ is dimensionless, expressed specifically in terms of the numbers of types of lattice constituents and their geometric properties. One now computes a transition amplitude $\mathcal{A}[\Gamma]$ in the quantum theory as the path sum
\begin{equation}\label{QFT4state}
\mathcal{A}[\Gamma]=\sum_{\substack{\mathcal{M} \\ \mathcal{M}|_{\partial\mathcal{M}}=\Gamma}}\mu_{(1,\tilde{L})}(\mathcal{M})\,e^{i\mathcal{S}_{\mathrm{cl}}[\mathcal{M}]/\hbar}.
\end{equation}
The spacetime $\Gamma$ constituting the boundary $\partial\mathcal{M}$ of a spacetime $\mathcal{M}$ specifies the transition amplitude $\mathcal{A}[\Gamma]$. The expression \eqref{QFT4state} is an instruction to sum over all spacetimes $\mathcal{M}$ satisfying the boundary condition $\mathcal{M}|_{\partial\mathcal{M}}=\Gamma$, weighting each spacetime $\mathcal{M}$ by the product of the measure $\mu_{(1,\tilde{L})}(\mathcal{M})$ and the exponential $e^{i\mathcal{S}_{\mathrm{cl}}[\mathcal{M}]/\hbar}$. The measure $\mu_{(1,\tilde{L})}(\mathcal{M})$ again indicates the presence of the lattice spacing serving as an ultraviolet regulator and the lattice extent serving as an infrared regulator. There is yet again some freedom in the definition of the measure $\mu_{(1,\tilde{L})}(\mathcal{M})$. At the very least one designs the measure $\mu_{(1,\tilde{L})}(\mathcal{M})$ to eliminate any redundancies of description of the theory $\mathscr{T}_{\mathrm{cl}}$ not already eliminated by the lattice regularization. Like in subsubsection \ref{QG} one cannot define the measure $\mu_{(1,\tilde{L})}(\mathcal{M})$ in terms of a mode decomposition of each spacetime $\mathcal{M}$; unlike in subsubsection \ref{QG} one does not encounter the problem of regularizing the measure since the measure $\mu_{(1,\tilde{L})}(\mathcal{M})$ is already regularized. 

As in subsubsection \ref{QFTLRNDST} the path summation in equation \eqref{QFT4state} often proves analytically intractable, so one resorts to numerical methods. To run Markov chain Monte Carlo simulations, one would like to Wick rotate each spacetime $\mathcal{M}$ from its Lorentzian to its Euclidean sector, transforming the complex weights $\mu(\mathcal{M})\,e^{i\mathcal{S}[\mathcal{M}]/\hbar}$ of the path sum into the real weights $\mu^{(\mathrm{E})}(\mathcal{M})\,e^{-\mathcal{S}^{(\mathrm{E})}[\mathcal{M}]/\hbar}$ of a partition function $\mathcal{Z}[\Gamma]$. Unfortunately, this Wick rotation is only well-defined for a certain subclass of spacetimes $\mathcal{M}$. Unless one restricts to this subclass (or a subclass of this subclass), as in the causal dynamical triangulations approach, one cannot employ Markov chain Monte Carlo methods. Since there is essentially no other known numerical method, one has limited recourse. I am concerned with the case of causal dynamical triangulations in paper II, so I now assume that one can perform a Wick rotation. A Markov chain Monte Carlo simulation then produces an ensemble of spacetimes $\mathcal{M}$ representative of those contributing to the partition function $\mathcal{Z}[\Gamma]$. By performing numerical measurements of discrete observables on this ensemble, one gleans information about the partition function $\mathcal{Z}[\Gamma]$ and the Lagrangians $\mathcal{L}_{\bar{\mathscr{T}}}^{(\mathrm{E})}[\mathcal{M}]$. One again estimates the expectation values of these discrete observables by their ensemble averages, mimicking the prescription of equation \eqref{ensembleaverage}. The implications of the physics contained within the partition function $\mathcal{Z}[\Gamma]$ for the physics contained within the path sum $\mathcal{A}[\Gamma]$ may again be far from clear unless an Osterwalder-Schrader-type theorem holds \cite{KO&RS1,KO&RS2}. 

\paragraph{Renormalization}

As I emphasized above, one in fact wishes to glean information about the Lagrangians $\mathscr{L}_{\mathscr{T}}^{(\mathrm{E})}[\mathbf{g}]$---and eventually their Lorentzian counterparts---providing the continuum description of the Lagrangians $\mathcal{L}_{\bar{\mathscr{T}}}^{(\mathrm{E})}[\mathcal{M}]$. Since the Lagrangians $\mathcal{L}_{\bar{\mathscr{T}}}^{(\mathrm{E})}[\mathcal{M}]$ are defined in the presence of a regularization, one should again be able to access the Lagrangians $\mathscr{L}_{\mathscr{T}}^{(\mathrm{E})}[\mathbf{g}]$ through a renormalization process. With the lattice spacing $a$ serving as an ultraviolet regulator and the lattice extent $L$ serving as an infrared regulator, one thus considers a limit in which the lattice spacing $a$ decreases to zero and the lattice extent $L$ increases without bound while physical quantities remain finite. Given only an ensemble of spacetimes $\mathcal{M}$ generated by Markov chain Monte Carlo simulations, one follows as closely as possible the renormalization group scheme explained in subsubsection \ref{QFTLRNDST}, making modifications as necessary to accommodate the difficulties encountered in subsubsection \ref{QG}. 

One first chooses a model for the continuum description, corresponding to some truncation of the action \eqref{metriceffectiveactionderexp}, and a compatible finite size scaling \emph{Ansatz}.\footnote{A model specified by the action \eqref{metriceffectiveactionderexp} might not suffice: the lattice regularization might result in a continuum description in which spacetime geometry is specified by more information than that contained within a symmetric metric tensor.} %under renormalization group transformations. % in accord with the chosen model always using the fact that renormalization group transformations do not result in changes in physics. 
%Suppose now that one wishes to compute the renormalization group flows of the couplings $c_{j}$ of the continuum limit for a renormalization group transformation that incrementally increases the ultraviolet scale $\tilde{\ell}_{\mathrm{UV}}$ and leaves fixed the infrared scale $\tilde{\ell}_{\mathrm{IR}}$. 
The model establishes the couplings whose renormalization group flows one wishes to extract and the physical observables through whose measurement one can infer the renormalized values of the couplings. One obtains the values of these physical observables by first performing numerical measurements of discrete observables on an ensemble of spacetimes $\mathcal{M}$ and then connecting these physical observables to the discrete observables \emph{via} the finite size scaling \emph{Ansatz}. %Together with the finite size scaling \emph{Ansatz}, the model also establishes those numerical measurements required. One extracts the renormalization group flows of the couplings $c_{j}$ by measuring appropriate discrete observables from which one can infer the values of $c_{j}$ via the finite size scaling \emph{Ansatz}. 
The model then allows one to delineate renormalization group trajectories as curves through the chosen truncation %$\mathfrak{t}$ of the space $\mathfrak{T}$ 
along which the physics of spacetime remains constant. 

One now requires a means to implement the renormalization group transformations that induce the flow along these trajectories. One may again employ a coarse graining scheme although I propose another approach in paper II. The construction of a suitable coarse graining scheme is now considerably more complicated. Its two operations---coarsening of the lattice regularization and coarsening of the field degrees of freedom---are inseparably intertwined since a lattice-regularized spacetime $\mathcal{M}$ itself encodes the field degrees of freedom. The structure of a typical lattice-regularized spacetime $\mathcal{M}$ is also irregular, requiring a blocking procedure adaptable to its local structure. Moreover, one is no longer licensed in assuming that any blocking procedure results in effective lattice units characterized by an effectively larger lattice spacing. Since all length scales are dynamically determined, a blocking operation can alter the dynamics resulting in effective lattice units characterized by an effectively smaller lattice spacing. If a blocking procedure preserves the physics encoded in an ensemble of spacetimes $\mathcal{M}$, then the blocking procedure necessarily has the effect of yielding an effectively larger lattice spacing since the physical scale of one lattice unit is indeed smaller than the physical scale of several lattice units including this one lattice unit.
%One check of whether or not a blocking procedure functions properly is thus to check that the effective lattice spacing is in fact dynamically larger. %One must devise a coarse graining scheme that both sums out small scale degrees of freedom and preserves physics on all larger scales.
How does one ascertain whether or not one's coarse graining scheme preserves this physics? There is again only one recourse: one must again monitor the values of physical observables for constancy under renormalization group transformations. 

This procedure calls for a standard unit of length in terms of which one measures the scales represented within expectation values of physical observables and the expectation values of dimensionful physical observables. As in subsubsection \ref{QG} within the data specified by the lattice-regularized spacetime $\Gamma$ constituting the boundary $\partial\mathcal{M}$, one may include information about a candidate standard unit of length. Since this boundary data is itself lattice-regularized, any standard unit of length encoded therein is not connected \emph{a priori} to any physical standard unit of length. The inclusion of appropriate boundary terms in the action $\mathcal{S}_{\mathrm{cl}}[\mathcal{M}]$, encoding the boundary data, again serves to establish the dynamical relationship between the scales defined by the boundary data and the scales emerging from the transition amplitude. This relationship endows the boundary data with physical meaning. One must ascertain whether or not this relationship ascribes the desired physical meaning to a candidate unit of length. Of course, if one works in a cosmological setting, however, then there are no fixed reference scales, and one must identify a bulk dimensionful physical observable to serve as a standard unit of length. % but this boundary data obtains its physical interpretation from the bulk data, so one cannot assume that a particular length in the boundary data is constant. (?)

Once one has devised a suitable coarse graining operation, one proceeds to extract the renormalization group flows of the couplings $c_{j}$. Starting from an ensemble of spacetimes $\mathcal{M}$ generated by Markov chain Monte Carlo methods, one iterates the renormalization group transformation on each spacetime $\mathcal{M}$. %(can only coarse grain, unlike in continuum description, so need to trace trajectories backwards) 
After each transformation one determines the renormalized values of the couplings $c_{j}$ by measuring the appropriate discrete observables and employing the finite size scaling \emph{Ansatz} to connect their values to the appropriate physical observables. One thus traces the renormalization group flows as the ultraviolet scale $\tilde{\ell}_{\mathrm{UV}}$ is incrementally increased, thereby inferring the Lagrangian $\mathscr{L}_{\mathscr{T}}^{(\mathrm{E})}[\mathbf{g}]$ at each step. %One extracts the renormalized couplings $c_{j}$ after each iteration of the renormalization group transformation as explained above. One thus constructs the renormalization group flows of the couplings $c_{j}$.

One again wishes to identify fixed points of the renormalization group transformation at which the ultraviolet regulator and the infrared regulator can be removed. One first searches for second order phase transitions of the partition function $\mathcal{Z}[\Gamma]$. If any second order phase transitions exist, then one searches for fixed points along these transitions by studying renormalization group trajectories in the vicinity of these transitions.

\subsection{Experimental observation of renormalization group flows}\label{experiment}

I wish finally to consider how one extracts the renormalization group flows of couplings from experimental observations. As I hope to demonstrate, this empirical process follows essentially the same procedure as that explained in subsubsection \ref{QGlat}. I take this as evidence for the cogency of that procedure. %explained in subsubsection \ref{QGlat}. % that the considerations discussed there are indeed relevant and necessary. 

Consider the measurement of the renormalization group flow of the fine structure constant or, equivalently, the electron charge $e$. As described, for instance, in \cite{L3}, the Bhabha scattering of electrons and positrons constitutes a phenomenon from which one can extract this renormalization group flow. %The method reported in \cite{L3} is the following. 
One prepares counterpropagating beams of electrons and of positrons, colliding them with a center of mass energy $E_{\mathrm{com}}$. One measures the scattering cross section $\Sigma_{\mathrm{obs}}(E_{\mathrm{com}})$ at the energy $E_{\mathrm{com}}$. One then incrementally increases the energy $E_{\mathrm{com}}$, again measuring the scattering cross section $\Sigma_{\mathrm{obs}}(E_{\mathrm{com}})$ after each increment. Quantum electrodynamics predicts the dependence of the Bhabha scattering cross section on the energy $E_{\mathrm{com}}$ and the electron charge $e$: $\Sigma_{\mathrm{QED}}(E_{\mathrm{com}},e)$. Assuming a scattering cross section of the form predicted by quantum electrodynamics, namely $\Sigma_{\mathrm{QED}}(E_{\mathrm{com}},e)$ and fitting this form to the measured scattering cross section $\Sigma_{\mathrm{obs}}(E_{\mathrm{com}})$ yields the observed electron charge $e_{\mathrm{obs}}(E_{\mathrm{com}})$ as a function of the energy $E_{\mathrm{com}}$. Quantum electrodynamics also predicts the renormalization group flow of the electron charge: $e_{\mathrm{QED}}(E_{\mathrm{com}})$. Given the electron charge $e_{0}$ at a reference (or matching) energy $E_{0}$, one compares $e_{\mathrm{obs}}(E_{\mathrm{com}})$ to $e_{\mathrm{QED}}(E_{\mathrm{com}})$, finding that the measurement of $e_{\mathrm{obs}}(E_{\mathrm{com}})$ agrees very well with the prediction of $e_{\mathrm{QED}}(E_{\mathrm{com}})$. 

Notice how closely this experimental procedure follows the procedure of subsubsection \ref{QGlat}. One first selects a model with which to analyze measurements, in this case quantum electrodynamics. (One clearly need not choose a finite size scaling \emph{Ansatz}.) One identifies a coupling within this model the renormalization group flow of which one hopes to measure, in this case the electron charge. One then identifies a physical observable---in this case the scattering cross section---the value of which yields the value of the coupling \emph{via} the model's prediction. One then makes measurements of the physical observable at different energy scales to obtain the scale dependence of this coupling. One directly generates ensembles of scattering events characterized by a range of center of mass energies $E_{\mathrm{com}}$. (One need not worry about deviating from a renormalization group trajectory because one always makes measurements in the same universe!) One determines by a statistical analysis how well these measurements agree with the model's predictions. Typically, one also interprets these measurements within the context of other models for the interactions of electrons and positrons, determining statistically which provides the best explanation of the data. 

How does one define the scale---in this case, the energy $E_{\mathrm{com}}$---at which the measurements are performed? One brings the scale of the measurements into contact with a suitably chosen standard unit of scale, itself determined dynamically. In particular, one measures the energy of the electrons and positrons in particular units, typically electron volts, which are based on certain dynamical standards, eventually tracing back to the definitions of the meter, second, and kilogram. (The standard unit is not influenced appreciably by the process of performing measurements.) The assignment of an energy to the electrons and positrons is based on the application of classical electrodynamics to the acceleration apparatus. This theory is a well-tested limit of one's model, quantum electrodynamics.

\section{A literature review}\label{review}

There is a small body of research on the renormalization of lattice-regularized quantum gravity models. Essentially all studies of such models fall within the context of three programs for the construction of quantum theories of gravity: quantum Regge calculus, Euclidean dynamical triangulations, and causal dynamical triangulations. %, both those obtained from lattice-regularization of a continuous model and those defined as discrete in the first place. 
I review this literature, assessing each study on the basis of the discussion of subsection \ref{concreteRG}, particularly subsubsection \ref{QGlat}. I confine my attention to studies in which the authors consider a renormalization group analysis, not just a phase structure analysis or a finite size scaling analysis. There is also a small body of research on the renormalization of other types of discrete quantum gravity models, specifically within the contexts of loop quantum gravity \cite{BB,BB&BD&FH&WK,BD,BD&FCE&MMB,BD&MMB&ES,BD&MMB&SS,BD&SS,ERL,ERL&DRT,FM1,FM2,RO}, group field theory and the tensor track \cite{JBG2,JBG1,JBG&ERL,JBG&VR,JBG&DOS,SC,SC&DO&VR1,SC&DO&VR2,AE&TK1,AE&TK2,LF&RG&DO,DOS&FVT,DOS}, and causal sets \cite{DPR&RDS}. Since these programs do not technically employ a lattice regularization, I do not review this literature here. 

\subsection{MacDowell-Mansouri formalism}

Working within the MacDowell-Mansouri formalism to allow for the application of Wilson's lattice gauge theoretic techniques \cite{SWM&FM,KGW}, Smolin defined a lattice-regularized quantum theory of gravity \cite{LS}. In the context of a weak coupling expansion, he reproduced the results of a standard perturbative quantization of Einstein gravity. In the context of a strong coupling expansion, he demonstrated that the dynamics becomes gapped and confining, and he introduced a Migdal-Kadanoff-type renormalization group transformation to provide evidence for asymptotic freedom. He did not, however, show this renormalization group transformation to be apt. On the basis of these results, Smolin argued for the presence of a phase transition possibly between the regimes probed by the two expansions.

\subsection{Quantum Regge calculus}

%Lewis \cite{SML}.

Hamber and Williams devised a renormalization group scheme based on a node decimation operation for $1$-dimensional quantum Regge calculus coupled to a scalar field \cite{HWH&RMW2}. They computed the quantum effective action resulting from one iteration of the decimation operation, %(in particular, the changes in the bare couplings). 
and they compared the result with that of an exact calculation, allowing for an assessment of the aptness of their coarse graining operation.

Martinelli and Marzuoli developed a renormalization group scheme for quantum Regge calculus employing a refinement operation based on cone subdivision \cite{MM&AM}. Making several simplifying assumptions, including that the quantum effective action after refinement has the same form as the action before refinement, they computed a condition on the iterative change in this action's sole coupling, the Newton constant. These authors did not address the aptness of their refinement operation.\footnote{Implemented on the lattice-regularized spacetimes of an ensemble generated by Markov chain Monte Carlo methods, a refinement operation cannot be apt---unless it is designed with complete knowledge of the physics contained in the partition function $\mathcal{Z}[\Gamma]$---for the following reason. These spacetimes do not contain any information concerning degrees of freedom on length scales smaller than that of the lattice spacing, so any such information introduced by the refinement operation does not reflect the physics contained in the partition function $\mathcal{Z}[\Gamma]$. Note that Martinelli and Marzuoli apply their refinement operation to the partition function $\mathcal{Z}[\Gamma]$ itself, not to an ensemble of lattice-regularized spacetimes representative of those contributing to the partition function $\mathcal{Z}[\Gamma]$.}

\subsection{Euclidean dynamical triangulations}

Renken introduced a renormalization group scheme for $2$-dimensional Euclidean dynamical triangulations \cite{RLR1}. He employed a blocking operation designed to preserve locally the graph geodesic distances between a triangulation's vertices. % based on a blocking operation that was designed to preserve geodesic distances between vertices. 
Since complete knowledge of the graph geodesic distances between vertices amounts to complete knowledge of the triangulation's geometry, his blocking operation plausibly preserves the physics encoded therein. He studied this renormalization group scheme with an Ising model coupled to the dynamical geometry, introducing a complementary blocking operation for the Ising model's spin degrees of freedom. To assess the aptness of both blocking operations, Renken monitored the values of several dimensionless discrete observables, both geometric and spinorial, under successive iterations of the blocking operations, apparently finding evidence for his scheme's aptness (in the limited sense of approximate preservation of these values). He also measured the critical value of the Ising model's coupling, finding agreement with known results. He did not attempt to connect any of these discrete observables to those of a model for the continuum description. He also did not attempt to verify directly that the blocking operation sums out small scale degrees of freedom although the reproduction of known results for the Ising model suggests that the blocking operation has this effect.

Renken, Catterall, and Kogut studied the first author's renormalization group scheme for $2$-dimensional Euclidean dynamical triangulations, now modified by the addition to the Lagrangian $\mathcal{L}_{\mathrm{cl}}^{(\mathrm{E})}[\mathcal{M}]$ of a particular class of perturbatively irrelevant higher order scalar functionals %believed to be irrelevant on the basis of perturbation theory 
\cite{RLR&SMC&JBK}. (These scalar functionals are irrelevant with respect to the known trivial fixed point of $2$-dimensional Euclidean dynamical triangulations \cite{JA}.) With each successive iteration of the blocking operation, they monitored the values of several dimensionless discrete observables to determine whether or not the operation preserved their values. Satisfied with this check of the operation's aptness, these authors then interpreted their results as providing evidence for the nonperturbative irrelevancy of the additional higher order scalar functionals. They did not attempt to connect any of the measured discrete observables to those of a model for the continuum description even though the continuum limit is known. % quantities do not have a clear analogue in the continuum limit and do not seem to test if the blocking is really cutting out small scale degrees of freedom. 

Thorleifsson and Catterall further investigated Renken's renormalization group scheme for $2$-dimensional Euclidean dynamical triangulations, now measuring instead the string susceptibility exponent \cite{GT&SC}. This quantity is a thoroughly studied physical observable of the continuum limit of $2$-dimensional Euclidean dynamical triangulations, namely quantum Liouville gravity \cite{JA}. These authors uncovered evidence for the inaptness of the blocking operation, leading them to implicate its failure to preserve graph geodesic distances beyond the local neighborhood of each vertex. %, but also globally might be the culprit, and they report preliminary evidence in favor of this idea. 
Thorleifsson and Catterall subsequently introduced a different renormalization group scheme for $2$-dimensional Euclidean dynamical triangulations \cite{GT&SC}. They chose to employ a node decimation operation designed to preserve locally the integrated scalar curvature. With each successive iteration of the decimation operation, they monitored the string susceptibility exponent and the Hausdorff dimension to determine whether or not the operation preserved their values. Like the string suspectibility exponent, the Hausdorff dimension is also a physical observable of quantum Liouville gravity \cite{JA}. They found good evidence for the preservation of these two physical observables under the action of their decimation operation. These authors also reconsidered the coupling to the Ising model and the addition of the same higher order scalar functionals. 

Gregory, Catterall, and Thorleifsson applied the last two authors' renormalization group scheme to $2$-dimensional Euclidean dynamical triangulations coupled to scalar fields \cite{EG&SMC&GT}. Measuring the string susceptibility exponent dressed by the scalar fields, these authors found further evidence for the aptness of the node decimation operation. Renken then generalized the node decimation operation to arbitrary dimensions \cite{RLR2,RLR3}. For $3$-dimensional and $4$-dimensional Euclidean dynamical triangulations, he followed the changes in two dimensionless discrete quantities---the number of $d$-simplices and the logarithm of the order of vertices---under successive coarse grainings, but he did not monitor any physical observables to check for their concurrent preservation. 

Johnston, Kownacki, and Krzywicki introduced yet another renormalization group scheme for $2$-\linebreak dimensional Euclidean dynamical triangulations \cite{DAJ&JPK&AK}. They proposed a blocking operation---fractal blocking or, more informally, (last generation) baby universe surgery---inspired by the hierarchical structure of typical $2$-dimensional Euclidean triangulations. Such a triangulation may be represented as a self-similar tree \cite{SJ&SDM}. This hierarchical structure largely organizes the geometrical degrees of freedom by scale, so the blocking operation quite conceivably sums out small scale degrees of freedom, and these authors provided preliminary evidence to this effect. Unfortunately, the fractal blocking scheme cannot be iterated as originally formulated, so one can only compare ensembles of singly blocked triangulations originating from ensembles of triangulations of different lattice spacetime $2$-volumes. 

Burda, Kownacki, and Krzywicki applied the last two authors' renormalization group scheme to $2$-dimensional and $4$-dimensional Euclidean dynamical triangulations \cite{ZB&JPK&AK}. They monitored the puncture-puncture correlation function---the average graph geodesic distance between two randomly chosen $d$-simplices\linebreak---which possesses a known continuous analogue (at least in two dimensions) \cite{JA}. From measurements of this discrete observable, they concluded that fractal blocking is apt. Employing a finite size scaling \emph{Ansatz} based on the spacetime $4$-volume, they claimed to find evidence for a stable ultraviolet fixed point in four dimensions. Bialas, Burda, Krzywicki, and Petersson applied baby universe surgery to $4$-dimensional Euclidean dynamical triangulations \cite{PB&ZB&AK&BP}. They monitored the puncture-puncture correlation function and the Hausdorff dimension. Although their renormalization group analysis supported the findings of Burda, Kownacki, and Krzywicki, specifically in finding evidence for a stable ultraviolet fixed point, a separate finite size scaling analysis demonstrated the associated phase transition to be of first order. Subsequent analyses have confirmed that the phase transition is indeed of first order \cite{BVdB,TR&PdF}, invalidating the findings of these renormalization group analyses. 

%There are several other papers \cite{} that address issues related to the renormalization of Euclidean dynamical triangulations but do not perform explicit renormalization group analyses.

\subsection{Causal dynamical triangulations}\label{CDTreview}

Henson developed a coarse graining scheme for causal dynamical triangulations based on the notion of a Delaunay triangulation \cite{JH}. Although he did not implement this scheme, he presented several arguments for its aptness. The output of one coarse graining operation is, however, not necessarily a causal triangulation. This fact jeopardizes the scheme's applicability for not only multiple iterations, but even one iteration since one may no longer be able to compute the same discrete observables after a coarse graining operation. 

Ambj{\o}rn, G\"{o}rlich, Jurkiewicz, Kreienbuehl, and Loll recently performed a first renormalization group analysis of $(3+1)$-dimensional causal dynamical triangulations. Quite similarly to the approach that I advocate in paper II, they did not employ a coarse graining scheme; instead, these authors proceeded as follows. They first generated a large number of ensembles of causal triangulations each characterized by different values of the bare couplings $\tilde{c}_{j}$ of the Lagrangian $\mathcal{L}_{\mathrm{cl}}^{(\mathrm{E})}[\mathcal{M}]$. % a wide swath of the phase space of phase C. 
By then studying the physics represented in each ensemble, as interpreted with respect to two different models for the continuum description, they organized these ensembles into renormalization group trajectories. In particular, employing a finite size scaling \emph{Ansatz} based on the spacetime $4$-volume, they identified two physical observables of a putative continuum description to define the constancy of physics along these trajectories. Instead of defining a standard unit of length on the basis of some dimensionful physical observable, they assumed that the lattice spacing has a fixed value (in some units) for all of the ensembles (at fixed number of $4$-simplices). Although this assumption allows them to employ the lattice spacing as a standard unit of length, the validity of this assumption is far from clear. For one choice of model, they found preliminary evidence for the existence of an ultraviolet fixed point along the known second order phase transition.

\section{Conclusion}\label{conclusion}

Building on extensive literature, I have developed a general procedure for the renormalization group analysis of a lattice-regularized quantum gravity model. %The main hurdle to overcome in this development was how to 
Making sense of the Wilsonian perspective on renormalization\linebreak---as chronicling the changes in a theory when probed on a succession of scales---in the absence of a fixed background structure providing for the definition of these scales---the context of nonperturbative approaches to the construction of quantum theories of gravity---proved the principle problem to resolve. %when one lacks a fixed background structure providing for the definition of these scales. 
The resolution is not profound---it merely required careful consideration of how one actually defines scales empirically. 

Necessarily, all scales originate in dynamical phenomena.\footnote{This statement is a tautology, but one should nevertheless keep it always in mind.} %, not in fixed structures. 
In the theoretical analysis of most experimental settings, one nevertheless presupposes the existence of a nondynamical background structure providing for the physically meaningful definition of scales measured in terms of a standard unit. This theoretical construct is a fixed spacetime, the geometry of which serves as a reference for the determination of scales. In the experimental implementation of these theoretical analyses, one attempts to assemble an approximation to such a structure from the physical resources at hand. This experimental realization is a spatiotemporal region, delimited by physical references, the geometry of which, measured by physical instruments calibrated to physical standards, is well-described as that of this fixed spacetime. %As such this structure is itself founded on dynamical phenomena, merely dynamical phenomena sufficiently decoupled from the dynamical phenomena that one's experiment aims to probe.  
The degree of precision of one's experiment dictates the degree to which the experimental realization need approximate the theoretical construct. 
%The level of approximation must prove sufficient for one's purposes: 
%Given the level of precision of one's experiment, 
Specifically, the experimental realization of the theoretical construct must act effectively as a nondynamical background structure, indiscernibly affected by the dynamical phenomena under investigation. As long as one can erect such a scaffolding, %such an approximation to the required nondynamical background structure,
one is licensed in supposing its existence for the purposes of a theoretical analysis. The structure so constructed is of course itself founded on dynamical phenomena, merely dynamical phenomena that are effectively nondynamical in comparison to and appropriately decoupled from the dynamical phenomena that one's experiment aims to probe. This decoupling is, however, not complete: one must correlate the dynamical phenomena that one's experiment aims to probe with the dynamical phenomena that one's experiment employs as background for the purpose of measuring the former's scales in terms of the latter's scales.

%The theoretical construct functioning as this nondynamical background structure is a fixed spacetime. Its experimental realization is a spatiotemporal region, delimited by physical references, the geometry of which, measured by physical clocks and rods, is well-described as that of this fixed spacetime. %, a topological manifold with geometry specified by a metric tensor. 
%This theoretical construct is a fixed background spacetime. Its experimental realization is a spacetime region that measurements by clocks and rods show to be a good approximation to this idealization. 

%One's experiment does not influence the conclusion that the spatiotemporal region in which it occurs is described as a particular fixed spacetime. 

%erects a scaffolding for the definition of scales that acts effectively as a background structure for the dynamical phenomena under investigation. This scaffolding is nevertheless founded on the basis of dynamical phenomena, simply dynamical phenomena sufficiently decoupled from that under investigation. %In analyzing such setting from a theoretical perspective, one is thus licensed in supposing the presence of a fixed background structure for the definition of scales. %the most other applications of renormalization, one is licensed in supposing an experimental setting in which the scaffolding for the physically meaningful definition of scales has already been erected, %assuming the presence of such a background structure in most applications of renormalization, 

In the context of nonperturbative approaches to the construction of quantum theories of gravity, one is precisely concerned with the dynamics of spacetime. The theoretical construct that one previously employed to provide for the definition of scales is thus no longer available. 
%If one cannot construct a structure providing for the definition of scales that is sufficiently decoupled from the dynamical phenomena under investigation, then one is no longer licensed in supposing its existence for the purposes of a theoretical analysis. 
%the scaffolding ceases to provide physically meaningful definitions of scales. 
%This situation obtains in the context of nonperturbative approaches to the construction of quantum theories of gravity. %, which revoke one's license. 
If one still requires a means for defining scales---for instance, for the purpose of performing a renormalization group analysis---then one must find an alternative method. Recalling how scales are established empirically---on the basis of dynamical phenomena---indicates how to proceed. Instead of presupposing the existence of a fixed background structure, one must erect a scaffolding for the definition of scales from within the quantum theory under consideration. In particular, one must identify physical observables whose dynamics establish scales, one of which possesses the %must identify a physical observable whose dynamics endow it with the 
properties of a standard unit of measure, and one must correlate these dynamically emergent scales with this standard unit of measure. %There is of course no guarantee that one can make this method work. %If one wants to perform an analysis calling for these structures, like a renormalization group analysis, then one must be able to construct these structures in some form. 
Above I have explained in broad generality how one puts this method to work for a renormalization group analysis of lattice-regularized quantum gravity models. 

The development of a general procedure stands quite apart from the specific application of the procedure. %It is in the application that the difficult work comes to the fore. 
Given a particular lattice-regularized quantum gravity model, one faces the twin tasks of determining precisely how to erect the scaffolding necessary to support a renormalization group analysis and how to implement the elements of this renormalization group analysis. Accomplishing these tasks requires %significant conceptual and technical insight. 
that one develop a conceptual and technical appreciation firstly for the physics contained within a candidate model for the continuum description and subsequently for how this physics is encoded within an ensemble of lattice-regularized spacetimes. Achieving such an understanding is the challenge of performing a renormalization group analysis of a lattice-regularized quantum gravity model. 

%To which discrete observables that possess physically meaningful analogues in the continuum description does one have numerical access? What physically meaningful observables can one deduce from one's model for the continuum description? Are these observables sufficient for extracting the renormalized values of the model's couplings? How do these discrete observables connect to their analogues in the continuum description? 

%Although one can in principle use a process of guess and check to arrive at a model and finite size scaling \emph{Ansatz}, one usually relies on conceptual insight into how discrete observables connect to continuous observables. 

%understand what discrete observables connect to their continuous counterparts %understand the dynamical emergence of scales. 
 
%One must develop a conceptual and technical appreciation for the physics contained within a candidate model for the continuum description sufficient to address these questions.  to be able to analyze a model so that one understands its physics.

%for any particular application, namely the identification of a model and a finite size scaling \emph{Ansatz} and a coarse graining scheme. 

I demonstrate in paper II how the renormalization group analysis of subsubsection \ref{QGlat} works in a concrete example, that of causal dynamical triangulations. I propose a particular renormalization group scheme in an attempt to determine whether or not this lattice-regularized quantum gravity model possesses a continuum limit. My proposed renormalization group scheme differs from that of Ambj{\o}rn, G\"{o}rlich, Jurkiewicz, Kreienbuehl, and Loll described in subsection \ref{CDTreview}, most notably in not assuming that one can assign the lattice spacing a value independently of the bare couplings characterizing an ensemble of causal triangulations. Hopefully, application of my proposed renormalization group scheme contributes to furthering the understanding of the causal dynamical triangulations program and its relation to the asymptotic safety program. 

Although this paper largely consists of a review of established knowledge, I hope that my insights into the renormalization group analysis of lattice-regularized quantum gravity models make a valuable addition to the literature.

\end{document}